\documentclass[aps,prb,amsmath,showpacs,superscriptaddress,twocolumn,floatfix]{revtex4}
\usepackage{graphicx}
\usepackage{color}

\begin{document}

\newcommand{\p}{^{\prime}}

\newcommand{\red}{\textcolor[rgb]{1.00,0.00,0.00}}
\newcommand{\green}{\textcolor[rgb]{0.00,1.00,0.00}}
\newcommand{\blue}{\textcolor[rgb]{0.00,0.00,1.00}}
\newcommand{\BaNiAs}{BaNi$_{\textnormal{2}}$As$_{\textnormal{2}} $\ }
\newcommand{\SrFeAs}{SrFe$_{\textnormal{2}}$As$_{\textnormal{2}} $\ }
\newcommand{\BaFeAs}{BaFe$_{\textnormal{2}}$As$_{\textnormal{2}} $\ }
\newcommand{\EuFeAs}{EuFe$_{\textnormal{2}}$As$_{\textnormal{2}} $\ }
\newcommand{\SrBaFeAs}{(Sr,Ba)Fe$_{\textnormal{2}}$As$_{\textnormal{2}}$\ }
\newcommand{\SrNiP}{SrNi$_{\textnormal{2}}$P$_{\textnormal{2}} $\ }
\newcommand{\SrKFeAs}{Sr$_{\textnormal{1-x}}$K$_{\textnormal{x}}$Fe$_{\textnormal{2}}$As$_{\textnormal{2}} $\ }
\newcommand{\half}{\frac{1}{2}}
\newcommand{\vecr}{{\textbf r}}
\newcommand{\CaFeAs}{CaFe$_{\textnormal{2}}$As$_{\textnormal{2}}$\ }
\newcommand{\BaKFeAs}{Ba$_{\textnormal{1-x}}$K$_{\textnormal{x}}$Fe$_{\textnormal{2}}$As$_{\textnormal{2}}$\ }
\newcommand{\BaFeCoAs}{Ba(Fe$_{\textnormal{1-x}}$Co$_{\textnormal{x}}$)$_{\textnormal{2}}$As$_{\textnormal{2}}$\ }
\newcommand{\LOFA}{LaOFeAs}
\newcommand{\tc}{$T_c$\ }
\newcommand{\sixteenth}{$\frac{1}{16}$}
\newcommand{\kvec}{\mathbf{k}}
\newcommand{\Kvec}{\mathbf{K}}
\newcommand{\Gsc}{\mathcal{G}}
\newcommand{\pipi}{($\pi$,$\pi$)\ }
\newcommand{\piz}{($\pi$,0)\ }
\def\k{\mathbf{k}}
\def\kp{\mathbf{k\p}}
\def\rr{\mathbf{r}}
\def\q{\mathbf{q}}
\def\wn{\omega_n}
\def\wnp{\omega_n^\prime}
\def\dxy{$d_{\textnormal{xy}}$\ }
\def\dxz{$d_{\textnormal{xz}}$\ }
\def\dyz{$d_{\textnormal{yz}}$\ }
\def\dzz{$d_{3\textnormal{z}^2-\textnormal{r}^2}$\ }
\def\dzzrr{$d_{3\textnormal{z}^2-\textnormal{r}^2}$\ }
\def\dxxyy{$d_{\textnormal{x}^2-\textnormal{y}^2}$\ }
\def\sixth{\frac{1}{6}}
\def\Jp{J^{\prime}}
\def\Up{U^{\prime}}
\def\imsigma{\Sigma^{\prime\prime}}
\newcommand{\Aoneg}{A$_{\mathrm{1g}}$ }
\newcommand{\Boneg}{B$_{\mathrm{1g}}$ }
\newcommand{\Btwog}{B$_{\mathrm{2g}}$ }

\title{Anisotropic quasiparticle lifetimes in Fe-based superconductors}

\author{A.F.~Kemper}
 \email[]{kemper@stanford.edu}
 \affiliation{Stanford Institute for Materials and Energy Science, SLAC National Accelerator Laboratory, Menlo Park, California 94025, USA}
 \affiliation{Geballe Laboratory for Advanced Materials, Stanford University, Stanford, California 94305, USA}
 \affiliation{Department of Physics, University of Florida, Gainesville, Florida 32611, USA}
\author{M.M.~Korshunov}
 \email[]{korshunov@phys.ufl.edu}
 \affiliation{Department of Physics, University of Florida, Gainesville, Florida 32611, USA}
 \affiliation{L.V. Kirensky Institute of Physics, Siberian Branch of Russian Academy of Sciences, 660036 Krasnoyarsk, Russia}
\author{T.P.~Devereaux}
 \affiliation{Stanford Institute for Materials and Energy Science, SLAC National Accelerator Laboratory, Menlo Park, California 94025, USA}
 \affiliation{Geballe Laboratory for Advanced Materials, Stanford University, Stanford, California 94305, USA}
\author{J.N.~Fry}
\affiliation{Department of Physics, University of Florida, Gainesville, Florida 32611, USA}
\author{H-P.~Cheng}
 \affiliation{Department of Physics, University of Florida, Gainesville, Florida 32611, USA}
\author{P.J.~Hirschfeld}
 \affiliation{Stanford Institute for Materials and Energy Science, SLAC National Accelerator Laboratory, Menlo Park, California 94025, USA}
 \affiliation{Geballe Laboratory for Advanced Materials, Stanford University, Stanford, California 94305, USA}
 \affiliation{Department of Physics, University of Florida, Gainesville, Florida 32611, USA}

\date{\today}

\begin{abstract}
We study the dynamical quasiparticle scattering by spin and charge fluctuations in Fe-based pnictides within a 5-orbital model with onsite interactions. The leading contribution to the scattering rate is calculated from the second-order diagrams with the polarization operator calculated in the random phase approximation. We find one-particle scattering rates which are highly anisotropic on each Fermi surface sheet due to the momentum dependence of the spin susceptibility and the multi-orbital composition of each Fermi pocket. This fact combined with the anisotropy of the effective mass, produce disparity between electrons and holes in conductivity, Hall coefficient, and Raman initial slope in qualitative agreement with experimental data.
\end{abstract}

\maketitle

\section{Introduction}
The presence of several electronic orbitals in bands near the Fermi level of a metallic system provides both a rich set of properties and complications in revealing the underlying physics. Some of the most widely discussed examples of such systems are the recently discovered Fe-based superconductors with $T_c$ up to 55K \cite{y_kamihara_08,zren08a} where multi-orbital effects cannot be disregarded. In these quasi-two-dimensional compounds, Fe $d$-orbitals form a Fermi surface (FS) consisting of nearly compensated small electron and hole pockets.\cite{s_lebegue_07,d_singh_08}
Since the sizes of the hole and electron FS pockets are roughly identical in the undoped system,
one might expect a vanishingly small Hall coefficient and a roughly electron-hole symmetric doping 
dependence.
However, in the intensively studied 122 systems (Ba(Fe$_{1-x}$Co$_{x})_2$As$_2$, Ba(Fe$_{1-x}$Ni$_{x})_2$As$_2$) and 1111 systems (LaFeAsO$_{1-x}$F$_{x}$ and SmFeAsO$_{1-x}$F$_{x}$), Hall effect measurements find that transport is dominated by the electrons even for the parent compounds\cite{f_rullier_albenque_09,l_fang_09,Kasahara_10,Liu_08,Riggs_09,Hess_09}. 
In the compensated case, this result can be explained only if the mobilities of holes and electrons are remarkably different which suggests an order of magnitude disparity in relaxation times, $\tau_e \gg \tau_h$. \cite{l_fang_09} A similar large asymmetry of electronic and hole scattering rates has also been suggested in the analysis of the electronic Raman measurements which can selectively probe different parts of the Brillouin zone (BZ) using various polarizations \cite{b_muschler_09}. 
Optical conductivity as measured by THz spectrometry provides and estimate of $\tau_e \approx 4\tau_h$, \cite{Maksimov_10}
{and reflectivity measurements also suggest the presence of two distinct scattering rates
with a large disparity between them.
\cite{n_barilic_10,j_tu_10,e_vanheumen_10}}
 Theoretical analysis of the normal state resistivity $\rho$ in the two-band model for Ba$_{1-x}$K$_x$Fe$_2$As$_2$ shows that the experimental temperature dependence $\rho(T)$ can be reproduced only if one assumes order of magnitude larger scattering in the hole band \cite{Golubov_10}. Finally, quantum oscillation experiments on P-doped systems indicate that the electron pockets have a longer mean free path \cite{j_analytis_10,a_coldea_08,j_analytis_09a}. It is clearly important to understand whether this conjectured dichotomy between electron and hole transport properties is real, and if it is universal to the Fe-based superconductors.

There are two main sources for quasiparticle decay: i) electron-electron inelastic processes and ii) impurity scattering. We will concentrate on the first case and mention impurity scattering only briefly. Experimentally, the apparent disparity in mobilities for holes and electrons becomes smaller as one dopes away from the magnetically ordered
parent compounds \cite{l_fang_09}. This suggests that the spin fluctuations which also decrease upon doping play an important role in the scattering rate asymmetry. Spin fluctuations due to the nearby spin-density wave (SDW) state have also been considered as the most probable source of superconducting pairing.\cite{i_mazin_08,k_kuroki_08,s_graser_08}

In this paper we study the inelastic quasiparticle scattering in Fe-based superconductors by calculating the scattering rate on different FS sheets  within the generalized spin-fluctuation theory. The self-energy is approximated via the second-order diagrams with the polarization operator treated in the random phase approximation (RPA). We show that there are two ingredients which provide strong anisotropy of the scattering rate.  

The most important one is that one-particle scattering is strongly affected by the orbital character of the initial and final states, in analogy to orbital pair scattering effects which have been discussed recently,\cite{t_maier_09b,a_kemper_10} leading to a  momentum dependence of the effective interaction. Secondly, the polarization bubble itself is momentum dependent. The combination results in a highly anisotropic scattering rate on the electron Fermi surface sheets, including some very long lived quasiparticle states. Although our results indicate that on the average $\tau_e$ is of the same order as $\tau_h$, the transport properties still may be dominated by small parts of the electron pockets where the lifetimes are long and the Fermi velocities are high.
This combination causes a disparity between holes and electrons in the transport properties (conductivity and Hall coefficient).  Furthermore, analysis of the Raman response shows that the quasiparticle lifetime effects can be clearly observed in both the \Boneg and \Btwog polarizations.

A calculation of the lifetime on Fermi surface was previously reported by
Onari et al.\cite{s_onari_10}, where the scattering due to spin fluctuations was considered within the
fluctuation-exchange approximation (FLEX). Our results reveal a similar momentum dependence of
the lifetimes, but exhibit a much larger anisotropy due to the absence of self-consistency.

\section{Model}
\begin{figure}
 \includegraphics[trim=0 17 0 0, clip, width=1.0\columnwidth]{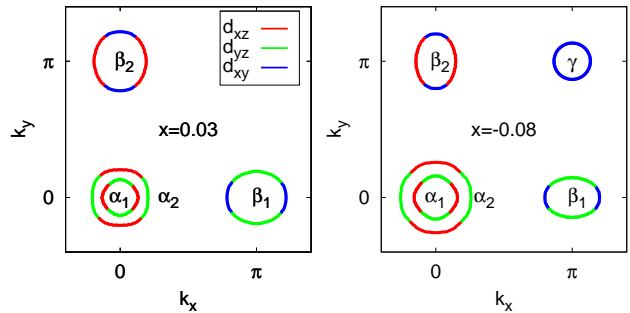}
 \caption{(Color online) Fermi surface for electron doped (doping $x=0.03$, left) and hole doped (doping $x=-0.08$, right) systems calculated within the 5-orbital model.}
\label{fig:FS}
\end{figure}

We will use the 5-orbital tight-binding model of Graser \textit{et al.}~\cite{s_graser_08} which is based on the \textit{ab initio} band structure calculations \cite{c_cao_08} within the local density approximation (LDA) for the prototypical iron pnictide, \LOFA. Our interaction Hamiltonian is
\begin{eqnarray}
H &=& H_{0} + U \sum_{i,m} n_{i m \uparrow} n_{i m \downarrow}   + U' \sum_{i,m<n} n_{i n} n_{i m} \nonumber\\
  &&+ J \sum_{i,m<n} \sum_{\sigma,\sigma'} c_{i n \sigma}^\dag c_{i m \sigma'}^\dag c_{i n \sigma'} c_{i m \sigma}
\nonumber\\
  &&+ J' \sum_{i, m \neq n} c_{i n \uparrow}^\dag c_{i n \downarrow}^\dag
  c_{i m \downarrow} c_{i m \uparrow},
\label{eq:H}
\end{eqnarray}
where $n_{i m} = n_{i m \uparrow} + n_{i m \downarrow}$, $n_{i m \sigma} = c_{i m \sigma}^\dag c_{i m \sigma}$, with $i$, $m$, and $\sigma$ denoting site, orbital, and spin indices, respectively.
The on-site intra- and inter-orbital Hubbard repulsions ($U$ and $U'$), Hund's rule coupling ($J$), and the pair hopping ($J'$) correspond to the notations of Kuroki \textit{et al.}~\cite{k_kuroki_08}
Below we will consider cases which obey spin-rotation invariance  {(SRI)}
through the relations $U'=U-2J$ and $J'=J$  {and those which do not}.
The kinetic energy $H_0$ includes the chemical potential $\mu$ and is described by a tight-binding model spanned by five Fe $d$-orbitals (\dxz, \dyz, \dxxyy, \dxy, \dzzrr) \cite{s_graser_08}. The \dxz, \dyz and \dxy bands dominate near Fermi level, as seen in Fig.~\ref{fig:FS} where we show the
Fermi surface (FS) which arises from $H_0$ in the one-Fe Brillouin zone.
For the electron- and undoped systems
the FS consists of two small hole pockets $\alpha_1$ and $\alpha_2$ around the $\Gamma=(0,0)$ point, and two small electron pockets $\beta_1$ and $\beta_2$ around the $X=(\pi,0)$ and $Y=(0,\pi)$ points, respectively.
Upon hole doping
a new hole FS pocket, $\gamma$, emerges around $(\pi,\pi)$ point, which has been shown to strongly affect the pairing state \cite{k_kuroki_09,a_kemper_10}.

\section{Method}
\begin{figure}
 \centering
 \includegraphics[width=0.99\columnwidth]{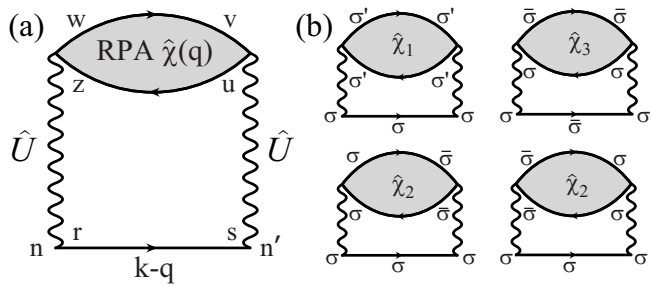}
 \caption{Orbital (a) and spin (b) structure of the second order diagram for the self-energy in the multi-orbital system, $\Sigma_{\bar{n}\bar{n}'}(\mathbf{k},\omega)$. Interaction lines contain four orbital indices, $\hat{U}=U_{nr}^{wz}$. Shaded bubble denote the RPA susceptibility, $\hat{\chi}(q)=\chi_{wz}^{vu}(\mathbf{q},\omega_q)$. Incoming and outgoing indices $\bar{n}$ and $\bar{n}'$ carry the same spin $\sigma$. $\hat{\chi}_1$, $\hat{\chi}_2$, and $\hat{\chi}_3$ are the different susceptibility channels, see Eq.~(\ref{eq:chi}), $\bar{\sigma}=-\sigma$.}
\label{fig:bubble_diagram}
\end{figure}

The leading non-vanishing contribution to the quasiparticle scattering rate
$1/\tau$
comes from the imaginary part of the second-order self-energy diagram
($\mathrm{Im}\ \Sigma$)
with the polarization bubble (see Fig.~\ref{fig:bubble_diagram}).
To take scattering from spin fluctuations into account we renormalize the bubble within the random phase approximation (RPA). Note that second order diagrams with crossing interaction lines are not included in Fig.~\ref{fig:bubble_diagram}.
We have chosen to work in this approximation to preserve consistency with calculations of the spin fluctuation pairing vertex \cite{s_graser_08}.
The bubble then represents the RPA susceptibility which in the multi-orbital system is $\chi_{wz}^{vu}(\mathbf{q},\omega_q)$ with $w,z,v,u$ being the orbital indices, and $\mathbf{q}$ and $\omega_q$ are the momentum and frequency, respectively. The same susceptibility was calculated
in Ref.~\onlinecite{s_graser_08} and was shown to produce superconductivity with
an A$_{1g}$
order parameter symmetry, in accord with several experiments\cite{j_paglione_10}
 and other spin fluctuation calculations \cite{s_graser_10, a_kemper_10, k_kuroki_08, k_kuroki_09}.
 Here and below the orbital (band) indices are denoted by Latin (Greek) letters.

Since we focus on the lifetime effects, we consider only $\mathrm{Im}\ \Sigma$, neglecting
the real part of the self-energy $\mathrm{Re}\ \Sigma$. The renormalization of the band structure due to the real part of the self-energy has been discussed in some detail in Refs.~\onlinecite{l_ortenzi_09,h_ikeda_10} and is not considered in the present study. We note that our calculations are based on the LDA band structure which already contains important Hartree corrections and agrees fairly well with quantum oscillation experiments \cite{j_analytis_10,a_coldea_08,j_analytis_09a}.

There are important consequences of the multi-orbital nature of the system which deserve comment. First,
the single-particle noninteracting Green function is diagonal in band space but not in orbital space.
The orbital matrix elements $a_{\mathbf{k}}^{n,\lambda}$, which describe the transformation from one space to another are given by $c_{\mathbf{k}n\sigma} = \sum\limits_{\lambda} a_{\mathbf{k}}^{n,\lambda} d_{\mathbf{k}\lambda\sigma}$, where $d_{\mathbf{k}\lambda\sigma}$ is the annihilation operator for a particle with band index $\lambda$, momentum $\mathbf{k}$ and energy $\varepsilon^\lambda_\mathbf{k}$.
Secondly, the interactions in Hamiltonian (\ref{eq:H}) have a complicated orbital structure; to compactify the expressions we define the local matrix interaction in orbital space, $U_{nr}^{wz} c_{i w \sigma_1}^\dag c_{i r \sigma_2}^\dag c_{i z \sigma_3} c_{i n \sigma_4}$, which accounts for all the quartic terms.

The noninteracting  part of the Hamiltonian, $H_0$, is a complex matrix \cite{s_graser_08} which in general has complex eigenvectors $a_{\mathbf{k}}^{n,\lambda}$, although the eigenvalues $\varepsilon^\lambda_\mathbf{k}$ are real. In order to use a simple form of the spectral representation of the Green function below, we choose a gauge in which the Hamiltonian is real by performing a unitary transformation
$\tilde{H_0} = \hat{\phi}^{-1} \hat{H}_0 \hat{\phi}$, where $\hat{\phi}$ is the diagonal matrix $\hat{\phi}=\mathrm{diag}\left(\mathrm{i},\mathrm{i},1,1,1\right)$.
The interaction part of the Hamiltonian (\ref{eq:H}) must then also be rotated by $\hat{\phi}$. Having completed the rotation, the eigenvectors and interactions are now real, and after calculating the diagram in Fig.~\ref{fig:bubble_diagram} we arrive at the multi-band extension of the standard zero-temperature expressions for the self-energy:
\begin{align}
\label{eq:imsigma}
 &\mathrm{Im}\ \Sigma_{\bar{n}\bar{n}'}(\mathbf{k},\omega) = \sum_{\mathbf{q},\lambda} \sum_{\bar{w},\bar{z},\bar{u},\bar{v},\bar{r},\bar{s}}
  U_{\bar{n}\bar{r}}^{\bar{w}\bar{z}} U_{\bar{s}\bar{n}'}^{\bar{u}\bar{v}} a^{r,\lambda}_{\mathbf{k}-\mathbf{q}} a^{s,\lambda}_{\mathbf{k}-\mathbf{q}} &\\
  &\times \mathrm{Im}\ \chi_{\bar{w}\bar{z}}^{\bar{v}\bar{u}}(\mathbf{q}, \omega - \varepsilon^\lambda_{\mathbf{k}-\mathbf{q}})
  \left[ \Theta\left(\varepsilon^\lambda_{\mathbf{k}-\mathbf{q}}\right) - \Theta\left(\varepsilon^\lambda_{\mathbf{k}-\mathbf{q}}-\omega\right) \right].&\nonumber
\end{align}
For simplicity, we have introduced the notation $\bar{s} = (s,\sigma_s)$, where $s$ and $\sigma_s$ are the orbital and spin index, respectively. The initial and final spins $\sigma_n$ and $\sigma_{n'}$, since we are considering the paramagnetic state, have been kept equal.

The momentum dependence of the orbital matrix elements generates an effective momentum-dependent interaction from the bare local Coulomb interactions,
\begin{equation}
\label{eq:V}
 V_{\bar{n},\lambda}^{\bar{w}\bar{z}}\left(\mathbf{k}-\mathbf{q} \right) = \sum\limits_{\bar{r}} U_{\bar{n}\bar{r}}^{\bar{w}\bar{z}} a^{r,\lambda}_{\mathbf{k}-\mathbf{q}},
\end{equation}
in terms of which (\ref{eq:imsigma}) may be written
\begin{align}
\label{eq:imsigmaVeff}
  &\mathrm{Im}\ \Sigma_{\bar{n}\bar{n}'}(\mathbf{k},\omega) =
\sum_{\mathbf{q},\lambda} \sum_{\bar{w},\bar{z},\bar{u},\bar{v}}
   V_{\bar{n},\lambda}^{\bar{w}\bar{z}}\left(\mathbf{q} \right)
   V_{\bar{n}',\lambda}^{\bar{v}\bar{u}}\left(\mathbf{q} \right) &\\
   &\times
\mathrm{Im}\ \chi_{\bar{w}\bar{z}}^{\bar{v}\bar{u}}(\mathbf{k}-\mathbf{q},
\omega - \varepsilon^\lambda_{\mathbf{q}})
   \left[ \Theta\left(\varepsilon^\lambda_{\mathbf{q}}\right) -
\Theta\left(\varepsilon^\lambda_{\mathbf{q}}-\omega\right)
\right].&\nonumber
\end{align}
The effective interaction enhances the anisotropy of the scattering rate, as will be demonstrated below.

We now discuss briefly the spin structure of the diagram in Fig.~\ref{fig:bubble_diagram} which is important for the calculation of $\mathrm{Im}\ \Sigma$ using Eq.~(\ref{eq:imsigma}). The susceptibility can be divided into charge and spin channels, and subsequently into singlet and triplet parts:
\begin{align}
\label{eq:chi}
 \chi_{\bar{w}\bar{z}}^{\bar{u}\bar{v}} &=
   \half (\chi^c)_{wz}^{uv} \delta_{\sigma_w\sigma_z} \delta_{\sigma_u\sigma_v} + \sixth (\chi^s)_{wz}^{uv} \vec\tau_{\sigma_w\sigma_z} \cdot \vec\tau_{\sigma_u\sigma_v} \nonumber\\
 & = \left\{
   \begin{array}{cc}
    \hat\chi_{1,2} \equiv \half (\chi^c)_{wz}^{uv} \pm \sixth (\chi^s)_{wz}^{uv} \quad & \mathrm{triplet} \\
    \hat\chi_{3} \equiv \frac{1}{3} (\chi^s)_{wz}^{uv} \quad &\mathrm{singlet}
   \end{array}
  \right.
\end{align}
where $\chi^c$ and $\chi^s$ are the charge and spin parts of the susceptibility, respectively, and $\vec\tau_{\sigma\sigma'}$ are Pauli spin matrices.

For the purpose of the self-energy calculation, the interactions can be grouped into three channels. If we denote the incoming spins as $\sigma_1$ and $\sigma_3$, and the outgoing as $\sigma_2$ and $\sigma_4$, the channels are:
(1) $\sigma_1 = \sigma_2 = \sigma_3 = \sigma_4$,
(2) $\sigma_1 = \sigma_2 \ne \sigma_3 = \sigma_4$,
(3) $\sigma_1 \ne \sigma_2 = \sigma_3 \ne \sigma_4$.
Then the orbital part of interactions in each channel, $\hat{U}_1$, $\hat{U}_2$, and $\hat{U}_3$, are:
\begin{align}
 \begin{array}{lll}
 (U_1)_{aa}^{aa} = 0 & (U_2)_{aa}^{aa} = U & (U_3)_{aa}^{aa} = -U \\
 (U_1)_{aa}^{bb} = \Up-J & (U_2)_{aa}^{bb} = \Up & (U_3)_{aa}^{bb} = -J \\
 (U_1)_{ab}^{ab} = 0 & (U_2)_{ab}^{ab} = \Jp &  (U_3)_{ab}^{ab} = -\Jp \\
 (U_1)_{ab}^{ba} = J-\Up & (U_2)_{ab}^{ba} = J &  (U_3)_{ab}^{ba} = -\Up
 \end{array}\nonumber
\end{align}
where orbital indices $a \ne b$.

To combine the interactions with the susceptibility, we first note that due to the spin structure of the diagram, the interaction channels (1)-(3) decouple. Second, we see by inspection that channels (1) and (2) couple to $\hat{\chi}_{1,2}$, and channel (3) couples to $\hat{\chi}_3$. Thus, the self-energy will contain the following matrix structure
\begin{equation}
 \hat{U} \hat{\chi} \hat{U} \propto \hat{U}_1 \hat{\chi}_1 \hat{U}_1 + \hat{U}_2 \hat{\chi}_1 \hat{U}_2 + \hat{U}_1 \hat{\chi}_2 \hat{U}_2 + \hat{U}_2 \hat{\chi}_2 \hat{U}_1 + \hat{U}_3 \hat{\chi}_3 \hat{U}_3.
\end{equation}
This expression by construction resolves the spin summation and only sums over orbital indices remain. 
Combining it with the calculation of $\chi_{wz}^{vu}(\mathbf{q},\omega_q)$ for a given doping
we use Eq.~\ref{eq:imsigma} to obtain $\mathrm{Im}\ \Sigma_{nn'}$ straightforwardly. Then we convert it to a band representation,
$\mathrm{Im}\ \Sigma_{\lambda\lambda'}(\mathbf{k},\omega) = \sum\limits_{n,n'} a^{n,\lambda}_{\mathbf{k}} \mathrm{Im}\ \Sigma_{nn'}(\mathbf{k},\omega) a^{n',\lambda'}_{\mathbf{k}}$. For the energy range where there are no band crossings, there is a unique band $\lambda$ corresponding to the momentum $\k$. The self-energy describes the scattering of the particle with $\k$ back to the same momentum $\k$, and thus back to the same band, $\lambda' = \lambda$. For the small energies around the Fermi level considered, there are no band crossings, so the major contribution to the scattering rate in the full Green function in band space, $\hat{G} = ( \hat{G}_0^{-1}-\hat{\Sigma} )^{-1}$, comes from diagonal, $\lambda=\lambda'$, matrix elements of $\mathrm{Im}\ \hat{\Sigma}$. We denote them as $\imsigma_\lambda(\mathbf{k},\omega) \equiv \mathrm{Im}\ \Sigma_{\lambda\lambda}(\mathbf{k},\omega)$.
The momentum sums in Eq.~\ref{eq:imsigmaVeff} were performed on a 256x256 grid with an
artificial broadening in all susceptibilities of 5 meV.
The undoped material has completely filled d$^6$ orbitals, which corresponds to $n_e=6$.
To present our results as a function of doping, we define it as $x=n_e-6$.

\section{Self-energy}
Because inter-band transitions are negligible in the range of energies considered here, the calculated scattering rate follows the Fermi liquid relation $\imsigma(\k,\omega) \propto \omega^2 + \pi^2 T^2$; thus, some finite frequency or temperature is needed for non-vanishing results. Here, and below, the quantities we report will be calculated at $\omega=20$meV which is equivalent to $T\approx74$K at zero frequency.  We have verified numerically that our results scale as $\omega^2$,
and that interband transitions indeed do not contribute at low energies.
The results below are qualitatively independent of the exact frequency chosen, since we are below
	the range of frequencies where inter-band scattering plays a large role.

\begin{figure*}[t!]
\includegraphics[width=0.96\textwidth]{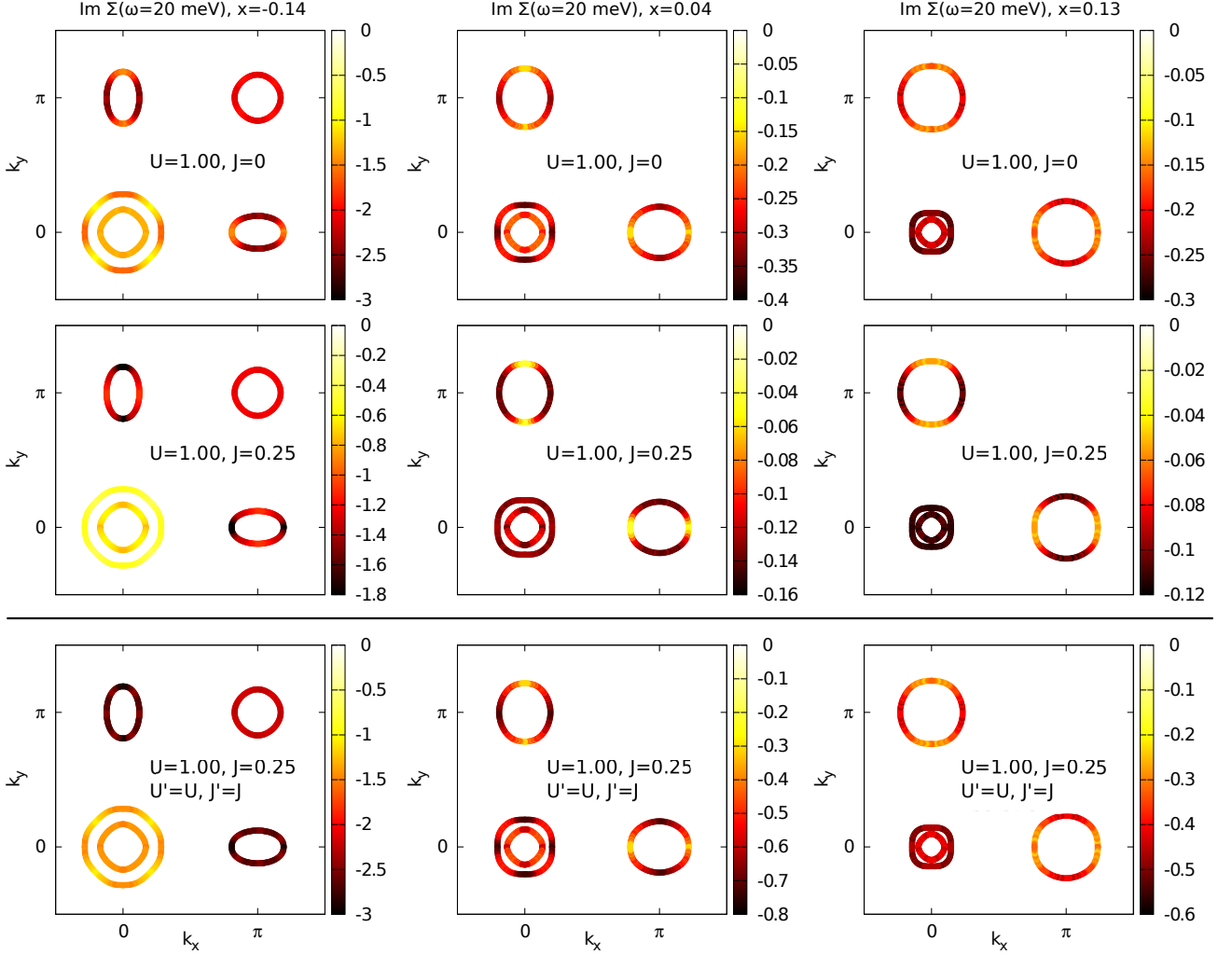}
 \caption{(Color online) Imaginary part of the self-energy $\Sigma$ at $\omega=20$meV along the Fermi surface for various dopings ($x=-0.14$, $0.04$, and $0.13$ from left to right) and for three sets of interaction parameters (in eV). All reported values are in meV. Note that the color scale is different for each plot. First and second row interaction parameters are SRI; third row is non-SRI.}
 \label{fig:imsigma}
\end{figure*}

For several dopings and few sets of interaction parameters, the calculated scattering rate along the Fermi surface is shown in Fig.~\ref{fig:imsigma}.
Here, $U$ and $J$ are in eV and were chosen to be close to the SDW-instability in the spin susceptibility.

\begin{figure}[h]
	\includegraphics[width=\columnwidth]{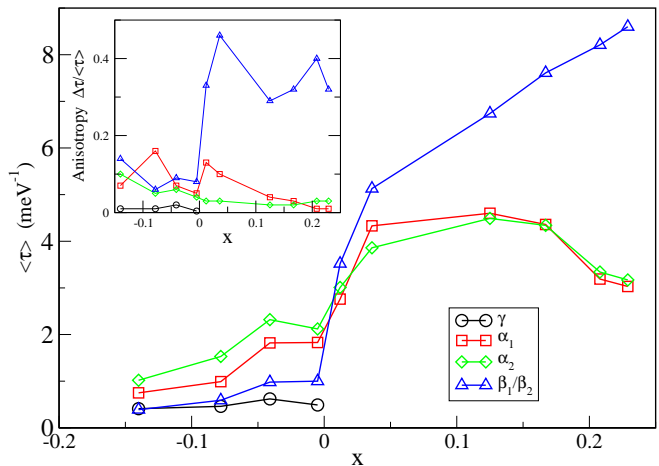}
	\caption{(Color online) Average scattering rate for holes ($\alpha_1$,$\alpha_2$,$\gamma$) and electrons ($\beta_1$,$\beta_2$) at $\omega=20$ meV and $U=1.0$ eV and $J=0.25$ eV. Inset: Lifetime anisotropy $\Delta\tau/\langle\tau\rangle$, where $\Delta\tau$
	($\langle\tau\rangle$) is the
	standard deviation (average) over the appropriate Fermi surface.}
	\label{fig:tau_ave}
\end{figure}

We observe that the average scattering rate increases monotonically with doping.
Fig.~\ref{fig:tau_ave} shows the average lifetime for holes and electrons on the Fermi surface,
as well as a measure of the anisotropy  which we have defined as the normalized
standard deviation of the lifetime,
$\Delta\tau/\langle\tau\rangle$, where $\tau_k = -1/2\Sigma''(k,\omega)$,
scaled by the average. We see a clear increase in the quasiparticle lifetime on all Fermi surface sheets
as the system is electron doped. On the electron-doped side, the average scattering rates
are essentially controlled by the degree of nesting. As more electrons are doped into the system,
the hole pockets shrink and the nesting between the $\alpha$ and $\beta$ sheets deteriorates.
The hole-doped systems have a smaller lifetime due to the presence of the $\gamma$ pocket;
in addition to $(\pi,0)$ scattering between $\alpha$ and $\beta$ sheets, new phase space for
scattering opens up and the average rate increases.
Since electrons on the \dxy portions of the FS are long-lived as will be discussed below,
one expects the resistivity due to spin-fluctuations to increase with hole doping. 

Aside from the overall change in scale, Fig.~\ref{fig:tau_ave}
shows that the ratio of electron to hole scattering rate changes as one goes from hole to electron
doping; electrons have a higher average scattering rate on the hole-doped side, and vice versa.
Although there is already an anisotropy between the hole and electron pockets in terms of lifetimes,
it is not enough to cause the experimentally observed anisotropy, as will be discussed below.
With electron doping, electron sheets $\beta_1$ and $\beta_2$
increase in size as well as $d_{xy}$ portions. Thus the number of states
with long lifetime for electrons increases monotonically.
On the other hand, hole pockets decrease in size and the phase space for
scattering decreases, while for small pockets the intraband scattering
starts to dominate. The competition of these two effects lead to saturation and then
to decrease of lifetime for holes, indicating that the intraband scattering dominates.

Next, 
we observe a clear anisotropy in the scattering rate going around the Fermi surfaces as
shown in Fig.~\ref{fig:imsigma} and the inset of Fig.~\ref{fig:tau_ave}.
Focusing first on the
undoped and electron-doped systems, the $\beta_1$ sheet exhibits strong anisotropy between the $\Gamma-X$ and $X-M$ directions. From Fig.~\ref{fig:FS}, we observe that this is where the Fermi surface orbital composition changes from \dxy to \dyz character.
There is a strong minimum in the scattering rate in the \dxy portions of the $\beta$ sheets;
this is due to the above-mentioned anisotropy of the effective interaction, Eq.~(\ref{eq:V}). The orbital matrix elements tend to restrict scattering to be maximal for intra-orbital processes. For the \dxy electrons, there is very little phase space to scatter compared to other orbitals, see Fig.~\ref{fig:FS}, because the spin fluctuation scattering intensity $\chi(\mathbf{q})$ is peaked at $\mathbf{q}=(\pi,0)$. Thus, they behave more like free electrons.
When the system is sufficiently hole doped to create the (\dxy) $\gamma$ hole pocket, \piz spin fluctuations couple them strongly to other \dxy states, causing the scattering rate there to increase.
Throughout the doping range, \dxz and \dyz states on the $\alpha$ pockets scatter strongly with their counterparts on the $\beta$ pockets, and vice versa.
 
Finally, we discuss the interaction dependence in Fig.~\ref{fig:imsigma}. The top row of panels shows a case where $J=0$, and the middle has finite $J=0.25$. 
As the Hund's rule coupling $J$ is turned on, we observe two effects. First, the overall scattering rate decreases (note that the color scale on each plot is different).  This is due to the spin-rotation invariance
(SRI) relation
$U'=U-2J$, so that $U'$ is decreased in the middle row of panels. Although new scattering channels open up through $J$ itself, this is more than compensated by the decrease in the inter-orbital scattering $U'$. This is  confirmed
by the third row in the figure, where $J$  {is finite but the system is non-SRI because} $U'=U$, as in the first row. Here, the scattering rate increases for all dopings, indicating that it is indeed the decrease in $U'$ that is the cause of the $\imsigma$ decrease in the 2nd row.

Secondly, we consider the effect of $J$ on the $\beta$ sheet anisotropy for the hole-doped system. When $J=0$, the minimum scattering rate occurs near the \dxy sections of the Fermi surfaces for all dopings.  
{Once} $J$ is turned on, the anisotropy reverses, and instead a maximum scattering rate is found on the same sections. This reversal of anisotropy can be explained by the same argument as above.  When $J=0$,
the intra-orbital and inter-orbital scattering ($U$ and $U'$) are the same. 
Thus, there is a strong scattering from both the \dxz/\dyz portions as well as the \dxy portions of the $\beta$ sheets to the $\gamma$ pocket (of \dxy character). Since the \dxz/\dyz portions additionally scatter to the $\alpha$ sheets, a stronger scattering rate occurs there. When $J$ is finite, the effective inter-orbital scattering rate  $U'$ decreases through the SRI relation. Thus, the scattering on the \dxz/\dyz  {portions is
decreased} while that on the \dxy sections remains the same.  {With sufficiently large $J$, the anisotropy on the $\beta$ sheets is reversed.} Note, however, that this argument depends on the existence of the $\gamma$ pocket. When the pocket is not present, such as in the undoped and electron doped cases, no such reversal occurs, and thus the \dxy states have the longest lifetimes for the configurations investigated.

\section{Comparison with experiment}

\subsection{Conductivity}

We next consider the effect of the calculated scattering rates on the electric conductivity. The total conductivity is the sum of the band conductivities, $\sigma(\omega) = \sum\limits_\lambda \sigma_{x \lambda}(\omega)$,
\begin{align}
 \sigma_{x \lambda}(\omega) = \frac{e^2}{\pi h} \int\limits_{\k \in \k_{F\lambda}} d\k N_\k v_{\k_x}^2 \tau_{\k}(\omega),
\label{eq:conductivity}
\end{align}
where $\tau_\k = -1/2\imsigma_\lambda(\k,\omega)$, $\k_{F\lambda}$ is the Fermi momentum for a particular band index $\lambda$, we integrate over $\k_\parallel$ which is the component of momentum along the FS, $v_{\k}$ is the velocity, and $N_{\k_{F\lambda}} = 1/|v_{\k_{F\lambda}}|$ is the momentum- and band-dependent density of states (DOS) at the Fermi level.  {Note that we have approximated the transport lifetime with the one-electron lifetime $\tau_\k$, neglecting forward scattering corrections, as well as the distinction between normal and Umklapp processes. Such an approximation can
only give the crude qualitative effect of the scattering from spin fluctuations on the conductivity.}

\begin{figure}[htpb]
 \includegraphics[angle=0,width=\columnwidth,clip=true]{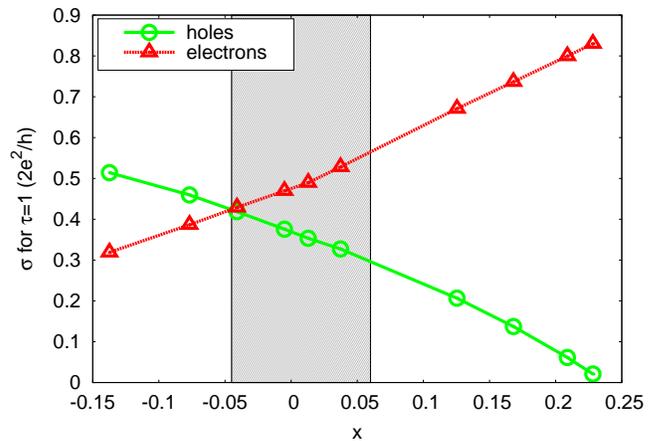}
 \includegraphics[angle=0,width=\columnwidth,clip=true]{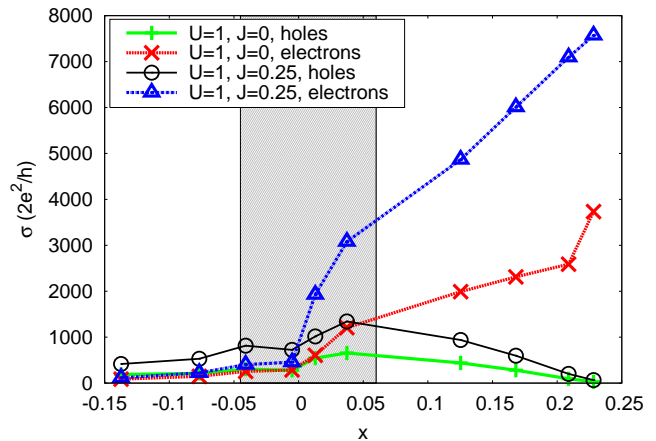}
 \caption{(Color online) Top: Conductivity for holes and electrons as a function of doping $x=n_e-6$ for constant relaxation rate $1/\tau=1$ eV. Bottom: conductivity for holes and electrons as a function of doping $x$ for the two sets of parameters (in eV): $U=1.0, J=0$ and $U=1.0, J=0.25$, at effective temperature $T=74K$. The shaded region marks the rough experimental SDW region in 122 systems. Solid lines are guides to the eye.}
\label{fig:conductivity}
\end{figure}

To analyze the doping-dependence of the conductivity, we now keep the interactions constant at values which do not produce an RPA instability over the range of dopings considered. We evaluate the DC conductivities at finite
temperature by replacing $1/\tau_\k(\omega)$ in Eq.~(\ref{eq:conductivity}) by $1/\tau_\k(\pi T)$. It is important to ask which aspects of the doping dependence of transport arise from purely kinematic effects
such as carrier density and Fermi velocity, which evolve with doping, and which arise from interactions. To illustrate this, we first plot in the top panel of Fig.~\ref{fig:conductivity} the separate contributions to the total conductivity from the electron and hole sheets, with an assumed constant relaxation time. Here the conductivities evolve more or less as expected with electron doping as the volumes of hole sheets shrink and electron sheets grow.
On the other hand, it is important that
  the ``perfectly compensated'' situation of equal kinetic conductivity of electrons and holes does not occur for the undoped case, but rather for $x \simeq -0.05$ hole doping. We have indicated in the figure the range of doping over which the 122 systems display long range magnetic order, which is not included in the current theory, and thus where the results are not directly applicable.

By contrast, the bottom panel of Fig.~\ref{fig:conductivity} shows the separate conductivities on the hole and electron FS as a function of doping. We immediately notice that conductivity for electrons grows quite strongly upon electron doping. 
Quite unlike the purely kinetic case in the top panel, the hole conductivity varies only weakly
compared to that of the electrons.
It is this asymmetry, due to a combination of the kinetic effects illustrated in the top panel of Fig. \ref{fig:conductivity} and lifetime effects calculated here, which lead to the rapid domination of the conductivity by electrons;
this has led transport experiments for Co-doped Ba-122 being interpreted in terms of a 1-band model with electrons only \cite{f_rullier_albenque_09,l_fang_09} with some validity.
The feature that greatly affects the doping dependence is the fact that the maximum of the Fermi velocity is precisely where the lifetime is largest on the electron FS sheets,
namely the \dxy sections of the $\beta$
sheets.
We also calculated  {conductivity and Hall coefficient} for a case where SRI is violated ($U=1.0$, $J=0.25$, $U'=U$, not shown in the figure). The result are  {qualitatively similar to the case where $U=1.0$, $J=0$.

The conductivities obtained show a large disparity between the hole- and electron-doped sides. It is
important to note that what we calculate here is the spin-fluctuation contribution to the scattering rate,
i.e. inelastic scattering.  Resistivity experiments on K-doped and Co-doped Ba122 show
that the elastic scattering is much larger in the Co-doped (e-doped) samples, presumably due to
the fact that the Co dopants sit in the FeAs plane.  This elastic scattering will correspondingly
reduce the e-doped side of Fig.\ref{fig:conductivity}, and thus bring the overall scattering rate
more in line with the experimentally observed trends.
We have not attempted to fit experiments directly due to
the current uncertainty in the details of the dopant scattering
potential.

The calculated conductivity shown in the lower panel of Fig.~\ref{fig:conductivity} was obtained for interaction parameters chosen sufficiently small to show the effect of doping while avoiding the RPA instability. For these parameters, the absolute scale of $\sigma$ is much larger than in experiments on 1111 or 122 samples. Clearly increasing the  {overall scale of the interactions}
will increase the scattering rates and decrease the conductivity. However to obtain the observed values of the conductivity requires approaching the RPA instability extremely closely. We have not attempted to fine tune the interaction strengths, but merely to illustrate the possible qualitative behavior. It seems more likely that a more complete theory will require a renormalization of the susceptibility akin to that seen in Quantum Monte Carlo (QMC) studies of the Hubbard model, which indicated that the RPA form of the dynamical magnetic response was qualitatively correct, but that the ``$U$'' driving the instability (through the RPA denominator) needed to be taken independent of the $U^2$ prefactor in the effective interaction \cite{t_maier_07}. A similar effect should occur in multi-orbital Hubbard models, such that the overall scales of scattering rates, and degree of proximity to the instability, should not be taken overly seriously.

\subsection{Hall coefficient.}

Any disparity between the scattering rates of electrons and holes manifests itself in the Hall coefficient
\begin{equation}
 R_H = - \sigma_{H}(\omega)/\sigma^2(\omega),
\label{eq:RH}
\end{equation}
where $\sigma_{H}(\omega)$ is the Hall conductivity \cite{Schulz_92,Kim_98}. For a multi-band system, $\sigma_{H}(\omega) = \sum\limits_{\lambda} \sigma_{H \lambda}(\omega)$ and the expression for the band Hall conductivity has the form
\begin{equation}
 \sigma_{H \lambda}(\omega) = \frac{e^3}{\pi h} \int\limits_{\k \in \k_{F\lambda}} d\k N_\k
 \mathbf{v}_{\k} \cdot \left[ \mathrm{Tr}(\mathbf{M}_{\k}^{-1}) - \mathbf{M}_{\k}^{-1} \right] \cdot \mathbf{v}_{\k}
 \tau_{\k}^2(\omega),
\label{eq:sigmaH}
\end{equation}
where $\left( \mathbf{M}_{\k}^{-1} \right)_{\alpha\beta} = \hbar^{-1} \partial v_{k_\alpha} / \partial k_\beta$ is the inverse mass tensor.

Fig.~\ref{fig:Hall} shows calculated $R_H$ as a function of doping for $\omega=20$meV (the corresponding effective temperature is 74K). One can qualitatively understand the doping dependence of $R_H$ by analyzing the approximate equation for the  {band} Hall conductivity,
\begin{equation}
 \sigma_{H \lambda}(\omega) \approx R_\lambda \sigma_{\lambda}^2(\omega).
\label{eq:sigmaHapprox}
\end{equation}
where $1/R_\lambda = \pm en_\lambda$ is the Hall coefficient for an electron (hole) band $\lambda$, and $n_\lambda$ is the occupation of that band. For the simple case of two bands (hole and electron) we have
\begin{equation}
 R_H^{2band} = \frac{1}{e} \frac{\sigma_h^2/n_h-\sigma_e^2/n_e} {\left(\sigma_h+\sigma_e\right)^2}.
\label{eq:RH2band}
\end{equation}
Since conductivity for the hole band $\sigma_h \propto n_h \tau_h / m_h$ and for the electron band $\sigma_e \propto n_e \tau_e / m_e$ with $\tau_{h,e}$ and $m_{h,e}$ being the corresponding lifetimes and band masses, $R_H^{2band}$ is a \textit{decreasing} function of electron doping if $\tau_e \sim \tau_h$ and $m_e \sim m_h$. This is what we see in Fig.~\ref{fig:Hall} for the $U=1.0, J=0$ case. On the other hand, experimental data for 1111 and 122 compounds indicate that $R_H^{expt}$ is an \textit{increasing} function of electron doping (i.e.,
the magnitude $|R_H^{expt}|$ decreases with increasing $x$) away from the SDW state. According to the simple analysis
 of Eq. (\ref{eq:RH2band}),
this may be due to (i) $\tau_e \gg \tau_h$ and/or (ii) $m_h \gg m_e$. Note that use of Eq.~(\ref{eq:sigmaH}) gives a different result from Eq.~(\ref{eq:sigmaHapprox}) due to the mass anisotropy across the FS which contributes to factor (ii). Factor (i) starts to play a role when we consider non-zero $J$. For the case of $U=1.0$ and $J=0.25$, $R_H(x)$ becomes slightly increasing function of $x$ for $x>0$ (Fig.~\ref{fig:Hall}).
However, it is not in quantitative agreement with experimental data. To see whether the present approach can provide the correct slope of $R_H(x)$, we artificially increased scattering rate on all orbitals except \dxy twice, so that the anisotropy between hole and electron sheets becomes more pronounced. The resulting doping dependence of the Hall coefficient is shown in Fig.~\ref{fig:Hall_artificial}. Now the slope of $R_H(x)$
is in good agreement with experimental data.

\begin{figure}
 \includegraphics[width=\columnwidth]{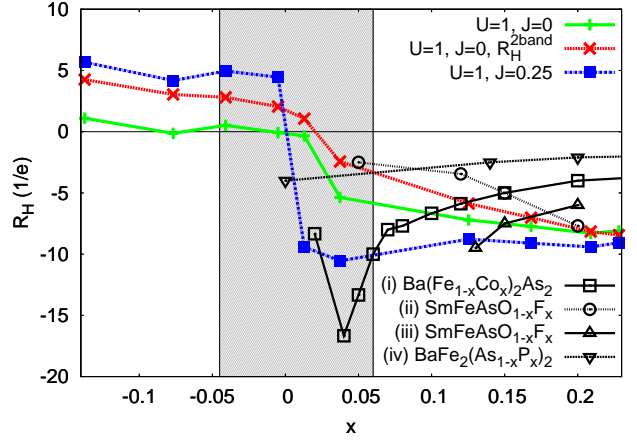}
 \caption{(Color online) Doping dependence of the Hall coefficient. The theoretical calculations are for two sets of parameters (in eV): $U=1.0$, $J=0$ and $U=1.0$, $J=0.25$. For the first set we also show result of the multi-band approximation for $R_H$ from Eq.~(\ref{eq:sigmaHapprox}). Experimental data points are from (i) Ref.~\onlinecite{l_fang_09} for Ba(Fe$_{1-x}$Co$_x$)$_2$As$_2$ at 100K, (ii) Ref.~\onlinecite{Riggs_09} and (iii) Ref.~\onlinecite{Liu_08} for SmFeAsO$_{1-x}$F$_x$ at 125K, and (iv) Ref.~\onlinecite{Kasahara_10} for BaFe$_2$(As$_{1-x}$P$_x$)$_2$ at 150K. The shaded region tentatively marks the experimental SDW region. Solid lines are guides for the eye.}
\label{fig:Hall}
\end{figure}
\begin{figure}
 \includegraphics[width=\columnwidth]{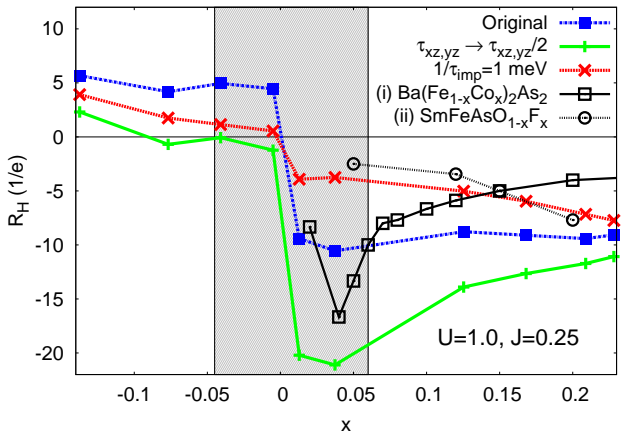}
 \caption{(Color online) Doping dependence of the Hall coefficient for three distinct cases: (1) original calculated $R_H$ from Fig.~\ref{fig:Hall}, (2) the one with the artificially increased scattering rate for all orbitals except for \dxy, $\tau_{xz,yz} \to \tau_{xz,yz}/2$, and (3) $R_H$ with added constant impurity scattering $1/\tau_{imp}=1$meV. For all cases parameters are $U=1.0$eV, $J=0.25$eV. Experimental data points (i) are from Ref.~\onlinecite{l_fang_09} for Ba(Fe$_{1-x}$Co$_x$)$_2$As$_2$ at 100K. The shaded region tentatively marks the experimental SDW region. Solid lines are guides for the eye.}
\label{fig:Hall_artificial}
\end{figure}

The fact that we underestimate the disparity between holes and electrons by a factor of two is not very discouraging. There are several factors not included in the present theory.
In the interest of studying the doping dependence, we have kept the interactions fairly low to avoid
the instability which occurs for relatively small interaction strengths on the hole-doped side.
Furthermore, we have neglected impurity scattering. In multi-band impurity models \cite{a_golubov_97,y_senga_08}, the ratio of intra- to inter-band scattering is taken as a parameter, and the scattering rate asymmetry between electrons and holes is weak. One might expect that an ``orbital impurity'' model, where an impurity introduces a local Coulomb potential for electrons in all $d$-orbitals, might produce a scattering rate anisotropy in $\k$-space due to the matrix elements $a_{\mathbf{k}}^{n,\lambda}$, just as in the inelastic scattering case. By investigating simple models similar to those considered in Ref.~\onlinecite{s_onari_09}, we have concluded that both average  {elastic} scattering rate asymmetry, and  {elastic} scattering rate anisotropy on any given Fermi surface sheet are small.
To address the effect of isotropic impurities on the Hall coefficient, we introduced a constant impurity scattering with a strength comparable to the calculated spin-fluctuation scattering rate $1/\tau_{\k}$. Since concurring scattering processes add to the self-energy,
the scattering rate is $1/\tau_{\k}^\mathrm{total} = 1/\tau_\mathrm{imp} + 1/\tau_{\k}$. Substituting $\tau_{\k}^\mathrm{total}$ in Eqs.~(\ref{eq:conductivity}) and (\ref{eq:sigmaH}), we find $R_H(x)$ shown in Fig.~\ref{fig:Hall_artificial} for $1/\tau_\mathrm{imp}=1$meV. 
Clearly, increasing disorder leads to a monotonically decreasing Hall coefficient with doping
similar to Eq.~\ref{eq:RH2band} with $\tau_e \simeq \tau_h$.
Thus dirtier samples 
will show a decrease of $R_H(x)$ with increasing electron doping.

The temperature dependence of $R_H$ deserves additional discussion. Recent phenomenological calculations of the self-energy in a two-band model for the pnictides  {suggest} that to reproduce experimentally observed $R_H(T)$ one needs to assume the non-Fermi liquid behavior of the spin susceptibility \cite{Prelovsek_10}. In particular, for large electron dopings, $R_H(T)$ is almost constant but for small $x$ it become an increasing function of temperature \cite{l_fang_09,Rullier-Albenque_10}. Here we argue that the observed temperature dependence can be qualitatively reproduced within our Fermi liquid approach. The resulting $R_H(T)$ from our calculations is shown in Fig.~\ref{fig:RH_T}. Note that the band which forms the $\gamma$ FS pocket for $x<0$ is slightly below the Fermi level for small positive $x$. Thus at finite energy or temperature the scattering to that band contributes to the self-energy and consequently to the transport properties. That is the main reason why $R_H(T)$ for $x=0.03$ is a rapidly changing function of $T$ in Fig.~\ref{fig:RH_T}.
\begin{figure}[htpb]
 \includegraphics[angle=0,width=\columnwidth,clip=true]{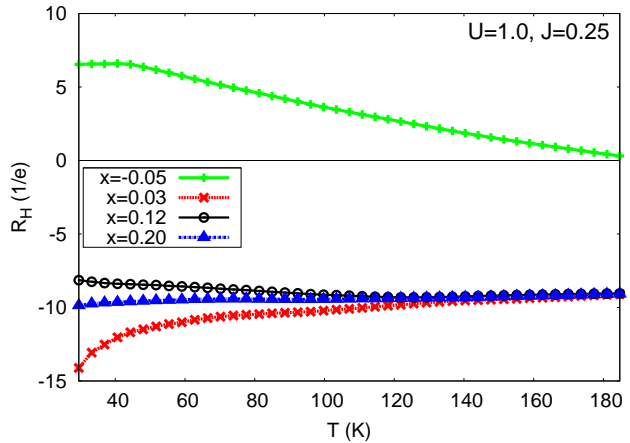}
 \caption{(Color online) Calculated dependence of the Hall coefficient on the effective temperature for several dopings $x$ with $U=1.0$eV, $J=0.25$eV.}
\label{fig:RH_T}
\end{figure}

\subsection{Raman response}

A momentum-sensitive probe of the scattering rate is provided by Raman spectroscopy. In particular, one can extract a scattering rate $\Gamma$ from Raman measurements by considering the slope of the Raman response in the limit as the energy loss $\Omega \rightarrow 0$.
\cite{t_devereaux_rmp} This quantity can be calculated as
\begin{align}
 \frac{1}{\Gamma_\gamma} =& \lim_{\Omega\rightarrow 0}  \frac{\partial \chi''_{\gamma\gamma}}{\partial \Omega} \nonumber \\
=&  \lim_{\Omega\rightarrow 0} N_F^{-1} \int\limits_{\k \in \k_F} d\k \frac{ N_\k \gamma_\k^2 }{ \imsigma(\k,\Omega) }
\label{eq:raman}
\end{align}
where $\gamma_\k$ denotes the Raman vertex related to the incident and
scattering polarizations (see e.g. Ref.~\onlinecite{t_devereaux_99}), and
$N_F$ is the density of states at the Fermi level.
Here, we have taken the simplest form for the Raman vertices allowed
by symmetry, namely $\cos(k_x) - \cos(k_y)$ 
and $\sin(k_x) \sin(k_y)$
for the \Boneg and \Btwog channels, respectively (note that we are
using the 1 Fe unit cell conventions).
We do not calculate the \Aoneg response due to the difficulties involved in calculating the
backflow effects.\cite{t_devereaux_96} In general, the backflow correction to the \Aoneg channel
involves the full susceptibility, not just the imaginary part.  Although this can in principle be
obtained, it is computationally expensive.

Fig.~\ref{fig:raman} shows the
lifetimes obtained from Raman scattering
according to the expression above.
\begin{figure}[t]
 \includegraphics[width=\columnwidth,clip=true]{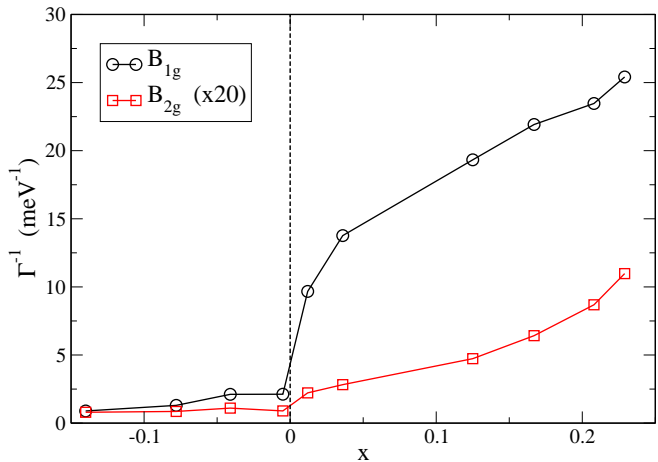}
 \caption{(Color online)  {Inverse of the} Raman scattering rate as function of doping for $U=1.0$ eV, $J=0.25$ eV. The \Btwog curve has been scaled by 20 for visibility.}
\label{fig:raman}
\end{figure}
As discussed in Muschler et al.~\cite{b_muschler_09,i_mazin_10a},
the \Boneg measurements probe the regions of the Brillouin zone containing the 
electron sheets. The \Btwog measurements probe the region around $(\pi/2,\pi/2)$, where there nominally
are no Fermi surfaces. This causes a decrease in the overall magnitude of the \Btwog Raman signal
compared to \Boneg,
as reflected in Fig.~\ref{fig:raman}.
However, the tails of the \Btwog Raman vertex extend out to the zone edges, and thus
some information can nevertheless be gleaned.  On the hole doped side, the hole pockets are large,
and the \Btwog vertex probes the edges of the hole pockets. Similarly, when the system is electron doped, the
electron pockets grow and the \Btwog vertex is thus larger there. The numerator of Eq.~\ref{eq:raman}
would give a symmetric doping dependence; therefore, 
the strong asymmetry is due to the lifetime effects.

We observe that the presence of the $\gamma$ pocket has a large effect in the Raman response,
for the same reasons as in the conductivity above. In particular the \Boneg signal shows a large
increase around zero doping. In the \Btwog channel the effect is not as strong, since there sections of
both hole and electron sheets are probed.

\section{Conclusions}

We have shown that the quasiparticle scattering due to spin-fluctuations in a multi-orbital model with local interactions can be significantly anisotropic. Two factors which produce this effect are the orbital matrix elements, which make interactions effectively momentum-dependent, and the momentum dependence of the dynamic susceptibility. In the particular case of our model for \LOFA, the \dxy portions of the electron FS experience little scattering due to the small scattering phase space in undoped and electron-doped cases, since there are no \dxy states on the hole sheets available for scattering. This anisotropy on the electron sheets appears to have profound consequences for transport in at least some Fe-pnictide systems.
We have noted that there are several factors which together determine the experimentally observed disparity between holes and electrons. The first is the longer lifetime of the \dxy states on the electron FS sheets. Another is the fact that the maximum of the Fermi velocity is precisely where the lifetime for electrons is largest.

Our calculations suggest that we underestimate slightly the asymmetry between \dxy and \dxz/\dyz states seen in the analysis of the Hall coefficient doping-dependence. We have discussed and critically analyzed factors which can provide additional anisotropy. Finally, we discussed aspects of the the electronic Raman scattering rate, and showed that the lifetime effects should be visible in both the \Boneg and \Btwog channels in
varying amounts.

\acknowledgments We thank O. Dolgov, R. Hackl, D. Maslov, I. Mazin, B. Muschler, V. Mishra, and J. Schmalian for useful discussions. 
A.F.K. and T.P.D. thank the Walther Mei\ss ner Institut for their hospitality.
A.F.K., M.M.K. and P.J.H. acknowledge support from DOE Grant, DE-FG02-05ER46236. M.M.K. acknowledges support from RFBR (Grant 09-02-00127), Presidium of RAS program ``Quantum physics of condensed matter'' N5.7, FCP Scientific and Research-and-Educational Personnel of Innovative Russia for 2009-2013 (GK P891), and President of Russia (Grant MK-1683.2010.2). A.F.K. and T.P.D. acknowledge support from DOE Grant DE-AC02-76SF00515.
H.-P. C. and J.N.F. acknowledge DOE/BES grant DE-FG02-02ER45995.

\bibliography{master}

\begin{thebibliography}{43}
\expandafter\ifx\csname natexlab\endcsname\relax\def\natexlab#1{#1}\fi
\expandafter\ifx\csname bibnamefont\endcsname\relax
  \def\bibnamefont#1{#1}\fi
\expandafter\ifx\csname bibfnamefont\endcsname\relax
  \def\bibfnamefont#1{#1}\fi
\expandafter\ifx\csname citenamefont\endcsname\relax
  \def\citenamefont#1{#1}\fi
\expandafter\ifx\csname url\endcsname\relax
  \def\url#1{\texttt{#1}}\fi
\expandafter\ifx\csname urlprefix\endcsname\relax\def\urlprefix{URL }\fi
\providecommand{\bibinfo}[2]{#2}
\providecommand{\eprint}[2][]{\url{#2}}

\bibitem[{\citenamefont{Kamihara et~al.}(2008)\citenamefont{Kamihara, Watanabe,
  Hirano, and Hosono}}]{y_kamihara_08}
\bibinfo{author}{\bibfnamefont{Y.}~\bibnamefont{Kamihara}},
  \bibinfo{author}{\bibfnamefont{T.}~\bibnamefont{Watanabe}},
  \bibinfo{author}{\bibfnamefont{M.}~\bibnamefont{Hirano}}, \bibnamefont{and}
  \bibinfo{author}{\bibfnamefont{H.}~\bibnamefont{Hosono}},
  \bibinfo{journal}{J. Am. Chem. Soc.} \textbf{\bibinfo{volume}{130}},
  \bibinfo{pages}{3296} (\bibinfo{year}{2008}).

\bibitem[{\citenamefont{{Zhi-An Ren, Wei Lu, Jie Yang, Wei Yi, Xiao-Li Shen,
  Zheng-Cai Li, Guang-Can Che, Xiao-Li Dong, Li-Ling Sun, Fang Zhou, Zhong-Xian
  Zhao}}(2008)}]{zren08a}
\bibinfo{author}{\bibnamefont{{Zhi-An Ren, Wei Lu, Jie Yang, Wei Yi, Xiao-Li
  Shen, Zheng-Cai Li, Guang-Can Che, Xiao-Li Dong, Li-Ling Sun, Fang Zhou,
  Zhong-Xian Zhao}}}, \bibinfo{journal}{Chin. Phys. Lett.}
  \textbf{\bibinfo{volume}{25}}, \bibinfo{pages}{2215} (\bibinfo{year}{2008}).

\bibitem[{\citenamefont{Leb\`egue}(2007)}]{s_lebegue_07}
\bibinfo{author}{\bibfnamefont{S.}~\bibnamefont{Leb\`egue}},
  \bibinfo{journal}{Phys. Rev. B} \textbf{\bibinfo{volume}{75}},
  \bibinfo{pages}{035110} (\bibinfo{year}{2007}).

\bibitem[{\citenamefont{Singh and Du}(2008)}]{d_singh_08}
\bibinfo{author}{\bibfnamefont{D.~J.} \bibnamefont{Singh}} \bibnamefont{and}
  \bibinfo{author}{\bibfnamefont{M.-H.} \bibnamefont{Du}},
  \bibinfo{journal}{Phys. Rev. Lett.} \textbf{\bibinfo{volume}{100}},
  \bibinfo{pages}{237003} (\bibinfo{year}{2008}).

\bibitem[{\citenamefont{Rullier-Albenque
  et~al.}(2009)\citenamefont{Rullier-Albenque, Colson, Forget, and
  Alloul}}]{f_rullier_albenque_09}
\bibinfo{author}{\bibfnamefont{F.}~\bibnamefont{Rullier-Albenque}},
  \bibinfo{author}{\bibfnamefont{D.}~\bibnamefont{Colson}},
  \bibinfo{author}{\bibfnamefont{A.}~\bibnamefont{Forget}}, \bibnamefont{and}
  \bibinfo{author}{\bibfnamefont{H.}~\bibnamefont{Alloul}},
  \bibinfo{journal}{Phys. Rev. Lett.} \textbf{\bibinfo{volume}{103}},
  \bibinfo{pages}{057001} (\bibinfo{year}{2009}).

\bibitem[{\citenamefont{Fang et~al.}(2009)\citenamefont{Fang, Luo, Cheng, Wang,
  Jia, Mu, Shen, Mazin, Shan, Ren et~al.}}]{l_fang_09}
\bibinfo{author}{\bibfnamefont{L.}~\bibnamefont{Fang}},
  \bibinfo{author}{\bibfnamefont{H.}~\bibnamefont{Luo}},
  \bibinfo{author}{\bibfnamefont{P.}~\bibnamefont{Cheng}},
  \bibinfo{author}{\bibfnamefont{Z.}~\bibnamefont{Wang}},
  \bibinfo{author}{\bibfnamefont{Y.}~\bibnamefont{Jia}},
  \bibinfo{author}{\bibfnamefont{G.}~\bibnamefont{Mu}},
  \bibinfo{author}{\bibfnamefont{B.}~\bibnamefont{Shen}},
  \bibinfo{author}{\bibfnamefont{I.~I.} \bibnamefont{Mazin}},
  \bibinfo{author}{\bibfnamefont{L.}~\bibnamefont{Shan}},
  \bibinfo{author}{\bibfnamefont{C.}~\bibnamefont{Ren}}, \bibnamefont{et~al.},
  \bibinfo{journal}{Phys. Rev. B} \textbf{\bibinfo{volume}{80}},
  \bibinfo{pages}{140508} (\bibinfo{year}{2009}).

\bibitem[{\citenamefont{Kasahara et~al.}(2010)\citenamefont{Kasahara,
  Shibauchi, Hashimoto, Ikada, Tonegawa, Okazaki, Shishido, Ikeda, Takeya,
  Hirata et~al.}}]{Kasahara_10}
\bibinfo{author}{\bibfnamefont{S.}~\bibnamefont{Kasahara}},
  \bibinfo{author}{\bibfnamefont{T.}~\bibnamefont{Shibauchi}},
  \bibinfo{author}{\bibfnamefont{K.}~\bibnamefont{Hashimoto}},
  \bibinfo{author}{\bibfnamefont{K.}~\bibnamefont{Ikada}},
  \bibinfo{author}{\bibfnamefont{S.}~\bibnamefont{Tonegawa}},
  \bibinfo{author}{\bibfnamefont{R.}~\bibnamefont{Okazaki}},
  \bibinfo{author}{\bibfnamefont{H.}~\bibnamefont{Shishido}},
  \bibinfo{author}{\bibfnamefont{H.}~\bibnamefont{Ikeda}},
  \bibinfo{author}{\bibfnamefont{H.}~\bibnamefont{Takeya}},
  \bibinfo{author}{\bibfnamefont{K.}~\bibnamefont{Hirata}},
  \bibnamefont{et~al.}, \bibinfo{journal}{Phys. Rev. B}
  \textbf{\bibinfo{volume}{81}}, \bibinfo{pages}{184519}
  (\bibinfo{year}{2010}).

\bibitem[{\citenamefont{Liu et~al.}(2008)\citenamefont{Liu, Wu, Wu, Fang, Chen,
  Li, Liu, Xie, Wang, Yang et~al.}}]{Liu_08}
\bibinfo{author}{\bibfnamefont{R.~H.} \bibnamefont{Liu}},
  \bibinfo{author}{\bibfnamefont{G.}~\bibnamefont{Wu}},
  \bibinfo{author}{\bibfnamefont{T.}~\bibnamefont{Wu}},
  \bibinfo{author}{\bibfnamefont{D.~F.} \bibnamefont{Fang}},
  \bibinfo{author}{\bibfnamefont{H.}~\bibnamefont{Chen}},
  \bibinfo{author}{\bibfnamefont{S.~Y.} \bibnamefont{Li}},
  \bibinfo{author}{\bibfnamefont{K.}~\bibnamefont{Liu}},
  \bibinfo{author}{\bibfnamefont{Y.~L.} \bibnamefont{Xie}},
  \bibinfo{author}{\bibfnamefont{X.~F.} \bibnamefont{Wang}},
  \bibinfo{author}{\bibfnamefont{R.~L.} \bibnamefont{Yang}},
  \bibnamefont{et~al.}, \bibinfo{journal}{Phys. Rev. Lett.}
  \textbf{\bibinfo{volume}{101}}, \bibinfo{pages}{087001}
  (\bibinfo{year}{2008}).

\bibitem[{\citenamefont{Riggs et~al.}(2009)\citenamefont{Riggs, McDonald,
  Kemper, Stegen, Boebinger, Balakirev, Kohama, Migliori, Chen, Liu
  et~al.}}]{Riggs_09}
\bibinfo{author}{\bibfnamefont{S.~C.} \bibnamefont{Riggs}},
  \bibinfo{author}{\bibfnamefont{R.~D.} \bibnamefont{McDonald}},
  \bibinfo{author}{\bibfnamefont{J.~B.} \bibnamefont{Kemper}},
  \bibinfo{author}{\bibfnamefont{Z.}~\bibnamefont{Stegen}},
  \bibinfo{author}{\bibfnamefont{G.~S.} \bibnamefont{Boebinger}},
  \bibinfo{author}{\bibfnamefont{F.~F.} \bibnamefont{Balakirev}},
  \bibinfo{author}{\bibfnamefont{Y.}~\bibnamefont{Kohama}},
  \bibinfo{author}{\bibfnamefont{A.}~\bibnamefont{Migliori}},
  \bibinfo{author}{\bibfnamefont{H.}~\bibnamefont{Chen}},
  \bibinfo{author}{\bibfnamefont{R.~H.} \bibnamefont{Liu}},
  \bibnamefont{et~al.}, \bibinfo{journal}{Journal of Physics: Condensed Matter}
  \textbf{\bibinfo{volume}{21}}, \bibinfo{pages}{412201}
  (\bibinfo{year}{2009}).

\bibitem[{\citenamefont{Hess et~al.}(2009)\citenamefont{Hess, Kondrat,
  Narduzzo, Hamann-Borrero, Klingeler, Werner, Behr, and Buchner}}]{Hess_09}
\bibinfo{author}{\bibfnamefont{C.}~\bibnamefont{Hess}},
  \bibinfo{author}{\bibfnamefont{A.}~\bibnamefont{Kondrat}},
  \bibinfo{author}{\bibfnamefont{A.}~\bibnamefont{Narduzzo}},
  \bibinfo{author}{\bibfnamefont{J.}~\bibnamefont{Hamann-Borrero}},
  \bibinfo{author}{\bibfnamefont{R.}~\bibnamefont{Klingeler}},
  \bibinfo{author}{\bibfnamefont{J.}~\bibnamefont{Werner}},
  \bibinfo{author}{\bibfnamefont{G.}~\bibnamefont{Behr}}, \bibnamefont{and}
  \bibinfo{author}{\bibfnamefont{B.}~\bibnamefont{Buchner}},
  \bibinfo{journal}{Europhys. Lett.} \textbf{\bibinfo{volume}{87}},
  \bibinfo{pages}{17005} (\bibinfo{year}{2009}).

\bibitem[{\citenamefont{Muschler et~al.}(2009)\citenamefont{Muschler, Prestel,
  Hackl, Devereaux, Analytis, Chu, and Fisher}}]{b_muschler_09}
\bibinfo{author}{\bibfnamefont{B.}~\bibnamefont{Muschler}},
  \bibinfo{author}{\bibfnamefont{W.}~\bibnamefont{Prestel}},
  \bibinfo{author}{\bibfnamefont{R.}~\bibnamefont{Hackl}},
  \bibinfo{author}{\bibfnamefont{T.~P.} \bibnamefont{Devereaux}},
  \bibinfo{author}{\bibfnamefont{J.~G.} \bibnamefont{Analytis}},
  \bibinfo{author}{\bibfnamefont{J.-H.} \bibnamefont{Chu}}, \bibnamefont{and}
  \bibinfo{author}{\bibfnamefont{I.~R.} \bibnamefont{Fisher}},
  \bibinfo{journal}{Phys. Rev. B} \textbf{\bibinfo{volume}{80}},
  \bibinfo{pages}{180510(R)} (\bibinfo{year}{2009}).

\bibitem[{\citenamefont{Maksimov et~al.}(2010)\citenamefont{Maksimov,
  Karakozov, Gorshunov, Nozdrin, Voronkov, Zhukova, Zhukov, Wu, Dressel, Haindl
  et~al.}}]{Maksimov_10}
\bibinfo{author}{\bibfnamefont{E.}~\bibnamefont{Maksimov}},
  \bibinfo{author}{\bibfnamefont{A.}~\bibnamefont{Karakozov}},
  \bibinfo{author}{\bibfnamefont{B.}~\bibnamefont{Gorshunov}},
  \bibinfo{author}{\bibfnamefont{V.}~\bibnamefont{Nozdrin}},
  \bibinfo{author}{\bibfnamefont{A.}~\bibnamefont{Voronkov}},
  \bibinfo{author}{\bibfnamefont{E.}~\bibnamefont{Zhukova}},
  \bibinfo{author}{\bibfnamefont{S.}~\bibnamefont{Zhukov}},
  \bibinfo{author}{\bibfnamefont{D.}~\bibnamefont{Wu}},
  \bibinfo{author}{\bibfnamefont{M.}~\bibnamefont{Dressel}},
  \bibinfo{author}{\bibfnamefont{S.}~\bibnamefont{Haindl}},
  \bibnamefont{et~al.}, \bibinfo{journal}{preprint}  (\bibinfo{year}{2010}),
  \eprint{arXiv:1008.3473}.

\bibitem[{\citenamefont{Barisic et~al.}(2010)\citenamefont{Barisic, Wu,
  Dressel, Li, Cao, and Xu}}]{n_barilic_10}
\bibinfo{author}{\bibfnamefont{N.}~\bibnamefont{Barisic}},
  \bibinfo{author}{\bibfnamefont{D.}~\bibnamefont{Wu}},
  \bibinfo{author}{\bibfnamefont{M.}~\bibnamefont{Dressel}},
  \bibinfo{author}{\bibfnamefont{L.~J.} \bibnamefont{Li}},
  \bibinfo{author}{\bibfnamefont{G.~H.} \bibnamefont{Cao}}, \bibnamefont{and}
  \bibinfo{author}{\bibfnamefont{Z.~A.} \bibnamefont{Xu}},
  \bibinfo{journal}{Phys. Rev. B} \textbf{\bibinfo{volume}{82}},
  \bibinfo{pages}{054518} (\bibinfo{year}{2010}).

\bibitem[{\citenamefont{{Tu} et~al.}(2010)\citenamefont{{Tu}, {Li}, {Liu},
  {Punnoose}, {Gong}, {Ren}, {Li}, {Cao}, {Xu}, and {Homes}}}]{j_tu_10}
\bibinfo{author}{\bibfnamefont{J.~J.} \bibnamefont{{Tu}}},
  \bibinfo{author}{\bibfnamefont{J.}~\bibnamefont{{Li}}},
  \bibinfo{author}{\bibfnamefont{W.}~\bibnamefont{{Liu}}},
  \bibinfo{author}{\bibfnamefont{A.}~\bibnamefont{{Punnoose}}},
  \bibinfo{author}{\bibfnamefont{Y.}~\bibnamefont{{Gong}}},
  \bibinfo{author}{\bibfnamefont{Y.~H.} \bibnamefont{{Ren}}},
  \bibinfo{author}{\bibfnamefont{L.~J.} \bibnamefont{{Li}}},
  \bibinfo{author}{\bibfnamefont{G.~H.} \bibnamefont{{Cao}}},
  \bibinfo{author}{\bibfnamefont{Z.~A.} \bibnamefont{{Xu}}}, \bibnamefont{and}
  \bibinfo{author}{\bibfnamefont{C.~C.} \bibnamefont{{Homes}}},
  \bibinfo{journal}{\prb} \textbf{\bibinfo{volume}{82}},
  \bibinfo{pages}{174509} (\bibinfo{year}{2010}).

\bibitem[{\citenamefont{{van Heumen} et~al.}(2010)\citenamefont{{van Heumen},
  {Huang}, {de Jong}, {Kuzmenko}, {Golden}, and {van der
  Marel}}}]{e_vanheumen_10}
\bibinfo{author}{\bibfnamefont{E.}~\bibnamefont{{van Heumen}}},
  \bibinfo{author}{\bibfnamefont{Y.}~\bibnamefont{{Huang}}},
  \bibinfo{author}{\bibfnamefont{S.}~\bibnamefont{{de Jong}}},
  \bibinfo{author}{\bibfnamefont{A.~B.} \bibnamefont{{Kuzmenko}}},
  \bibinfo{author}{\bibfnamefont{M.~S.} \bibnamefont{{Golden}}},
  \bibnamefont{and} \bibinfo{author}{\bibfnamefont{D.}~\bibnamefont{{van der
  Marel}}}, \bibinfo{journal}{Europhysics Letters}
  \textbf{\bibinfo{volume}{90}}, \bibinfo{pages}{37005} (\bibinfo{year}{2010}).

\bibitem[{\citenamefont{Golubov et~al.}(2010)\citenamefont{Golubov, Dolgov,
  Boris, Charnukha, Sun, Lin, and Shevchun}}]{Golubov_10}
\bibinfo{author}{\bibfnamefont{A.}~\bibnamefont{Golubov}},
  \bibinfo{author}{\bibfnamefont{O.}~\bibnamefont{Dolgov}},
  \bibinfo{author}{\bibfnamefont{A.}~\bibnamefont{Boris}},
  \bibinfo{author}{\bibfnamefont{A.}~\bibnamefont{Charnukha}},
  \bibinfo{author}{\bibfnamefont{D.}~\bibnamefont{Sun}},
  \bibinfo{author}{\bibfnamefont{C.}~\bibnamefont{Lin}}, \bibnamefont{and}
  \bibinfo{author}{\bibfnamefont{A.}~\bibnamefont{Shevchun}},
  \bibinfo{journal}{preprint}  (\bibinfo{year}{2010}),
  \eprint{arXiv:1011.1900}.

\bibitem[{\citenamefont{Analytis et~al.}(2010)\citenamefont{Analytis, Chu,
  McDonald, Riggs, and Fisher}}]{j_analytis_10}
\bibinfo{author}{\bibfnamefont{J.~G.} \bibnamefont{Analytis}},
  \bibinfo{author}{\bibfnamefont{J.-H.} \bibnamefont{Chu}},
  \bibinfo{author}{\bibfnamefont{R.~D.} \bibnamefont{McDonald}},
  \bibinfo{author}{\bibfnamefont{S.~C.} \bibnamefont{Riggs}}, \bibnamefont{and}
  \bibinfo{author}{\bibfnamefont{I.~R.} \bibnamefont{Fisher}},
  \bibinfo{journal}{Phys. Rev. Lett.} \textbf{\bibinfo{volume}{105}},
  \bibinfo{pages}{207004} (\bibinfo{year}{2010}).

\bibitem[{\citenamefont{Coldea et~al.}(2008)\citenamefont{Coldea, Fletcher,
  Carrington, Analytis, Bangura, Chu, Erickson, Fisher, Hussey, and
  McDonald}}]{a_coldea_08}
\bibinfo{author}{\bibfnamefont{A.~I.} \bibnamefont{Coldea}},
  \bibinfo{author}{\bibfnamefont{J.~D.} \bibnamefont{Fletcher}},
  \bibinfo{author}{\bibfnamefont{A.}~\bibnamefont{Carrington}},
  \bibinfo{author}{\bibfnamefont{J.~G.} \bibnamefont{Analytis}},
  \bibinfo{author}{\bibfnamefont{A.~F.} \bibnamefont{Bangura}},
  \bibinfo{author}{\bibfnamefont{J.-H.} \bibnamefont{Chu}},
  \bibinfo{author}{\bibfnamefont{A.~S.} \bibnamefont{Erickson}},
  \bibinfo{author}{\bibfnamefont{I.~R.} \bibnamefont{Fisher}},
  \bibinfo{author}{\bibfnamefont{N.~E.} \bibnamefont{Hussey}},
  \bibnamefont{and} \bibinfo{author}{\bibfnamefont{R.~D.}
  \bibnamefont{McDonald}}, \bibinfo{journal}{Phys. Rev. Lett.}
  \textbf{\bibinfo{volume}{101}}, \bibinfo{pages}{216402}
  (\bibinfo{year}{2008}).

\bibitem[{\citenamefont{Analytis et~al.}(2009)\citenamefont{Analytis, Andrew,
  Coldea, McCollam, Chu, McDonald, Fisher, and Carrington}}]{j_analytis_09a}
\bibinfo{author}{\bibfnamefont{J.~G.} \bibnamefont{Analytis}},
  \bibinfo{author}{\bibfnamefont{C.~M.~J.} \bibnamefont{Andrew}},
  \bibinfo{author}{\bibfnamefont{A.~I.} \bibnamefont{Coldea}},
  \bibinfo{author}{\bibfnamefont{A.}~\bibnamefont{McCollam}},
  \bibinfo{author}{\bibfnamefont{J.-H.} \bibnamefont{Chu}},
  \bibinfo{author}{\bibfnamefont{R.~D.} \bibnamefont{McDonald}},
  \bibinfo{author}{\bibfnamefont{I.~R.} \bibnamefont{Fisher}},
  \bibnamefont{and}
  \bibinfo{author}{\bibfnamefont{A.}~\bibnamefont{Carrington}},
  \bibinfo{journal}{Phys. Rev. Lett.} \textbf{\bibinfo{volume}{103}},
  \bibinfo{pages}{076401} (\bibinfo{year}{2009}).

\bibitem[{\citenamefont{Mazin et~al.}(2008)\citenamefont{Mazin, Singh,
  Johannes, and Du}}]{i_mazin_08}
\bibinfo{author}{\bibfnamefont{I.~I.} \bibnamefont{Mazin}},
  \bibinfo{author}{\bibfnamefont{D.~J.} \bibnamefont{Singh}},
  \bibinfo{author}{\bibfnamefont{M.~D.} \bibnamefont{Johannes}},
  \bibnamefont{and} \bibinfo{author}{\bibfnamefont{M.~H.} \bibnamefont{Du}},
  \bibinfo{journal}{Phys. Rev. Lett.} \textbf{\bibinfo{volume}{101}},
  \bibinfo{pages}{057003} (\bibinfo{year}{2008}).

\bibitem[{\citenamefont{Kuroki et~al.}(2008)\citenamefont{Kuroki, Onari, Arita,
  Usui, Tanaka, Kontani, and Aoki}}]{k_kuroki_08}
\bibinfo{author}{\bibfnamefont{K.}~\bibnamefont{Kuroki}},
  \bibinfo{author}{\bibfnamefont{S.}~\bibnamefont{Onari}},
  \bibinfo{author}{\bibfnamefont{R.}~\bibnamefont{Arita}},
  \bibinfo{author}{\bibfnamefont{H.}~\bibnamefont{Usui}},
  \bibinfo{author}{\bibfnamefont{Y.}~\bibnamefont{Tanaka}},
  \bibinfo{author}{\bibfnamefont{H.}~\bibnamefont{Kontani}}, \bibnamefont{and}
  \bibinfo{author}{\bibfnamefont{H.}~\bibnamefont{Aoki}},
  \bibinfo{journal}{Phys. Rev. Lett.} \textbf{\bibinfo{volume}{101}},
  \bibinfo{pages}{087004} (\bibinfo{year}{2008}).

\bibitem[{\citenamefont{Graser et~al.}(2009)\citenamefont{Graser, Maier,
  Hirschfeld, and Scalapino}}]{s_graser_08}
\bibinfo{author}{\bibfnamefont{S.}~\bibnamefont{Graser}},
  \bibinfo{author}{\bibfnamefont{T.~A.} \bibnamefont{Maier}},
  \bibinfo{author}{\bibfnamefont{P.~J.} \bibnamefont{Hirschfeld}},
  \bibnamefont{and} \bibinfo{author}{\bibfnamefont{D.~J.}
  \bibnamefont{Scalapino}}, \bibinfo{journal}{New. J. Phys.}
  \textbf{\bibinfo{volume}{11}}, \bibinfo{pages}{025016}
  (\bibinfo{year}{2009}).

\bibitem[{\citenamefont{Maier et~al.}(2009)\citenamefont{Maier, Graser,
  Scalapino, and Hirschfeld}}]{t_maier_09b}
\bibinfo{author}{\bibfnamefont{T.~A.} \bibnamefont{Maier}},
  \bibinfo{author}{\bibfnamefont{S.}~\bibnamefont{Graser}},
  \bibinfo{author}{\bibfnamefont{D.~J.} \bibnamefont{Scalapino}},
  \bibnamefont{and} \bibinfo{author}{\bibfnamefont{P.~J.}
  \bibnamefont{Hirschfeld}}, \bibinfo{journal}{Phys. Rev. B}
  \textbf{\bibinfo{volume}{79}}, \bibinfo{pages}{224510}
  (\bibinfo{year}{2009}).

\bibitem[{\citenamefont{{Kemper} et~al.}(2010)\citenamefont{{Kemper}, {Maier},
  {Graser}, {Cheng}, {Hirschfeld}, and {Scalapino}}}]{a_kemper_10}
\bibinfo{author}{\bibfnamefont{A.~F.} \bibnamefont{{Kemper}}},
  \bibinfo{author}{\bibfnamefont{T.~A.} \bibnamefont{{Maier}}},
  \bibinfo{author}{\bibfnamefont{S.}~\bibnamefont{{Graser}}},
  \bibinfo{author}{\bibfnamefont{H.}~\bibnamefont{{Cheng}}},
  \bibinfo{author}{\bibfnamefont{P.~J.} \bibnamefont{{Hirschfeld}}},
  \bibnamefont{and} \bibinfo{author}{\bibfnamefont{D.~J.}
  \bibnamefont{{Scalapino}}}, \bibinfo{journal}{New Journal of Physics}
  \textbf{\bibinfo{volume}{12}}, \bibinfo{pages}{073030}
  (\bibinfo{year}{2010}).

\bibitem[{\citenamefont{{Onari} and {Kontani}}(2010)}]{s_onari_10}
\bibinfo{author}{\bibfnamefont{S.}~\bibnamefont{{Onari}}} \bibnamefont{and}
  \bibinfo{author}{\bibfnamefont{H.}~\bibnamefont{{Kontani}}}
  (\bibinfo{year}{2010}), \eprint{arXiv:1009.3882}.

\bibitem[{\citenamefont{Cao et~al.}(2008)\citenamefont{Cao, Hirschfeld, and
  Cheng}}]{c_cao_08}
\bibinfo{author}{\bibfnamefont{C.}~\bibnamefont{Cao}},
  \bibinfo{author}{\bibfnamefont{P.~J.} \bibnamefont{Hirschfeld}},
  \bibnamefont{and} \bibinfo{author}{\bibfnamefont{H.-P.} \bibnamefont{Cheng}},
  \bibinfo{journal}{Phys. Rev. B} \textbf{\bibinfo{volume}{77}},
  \bibinfo{pages}{220506(R)} (\bibinfo{year}{2008}).

\bibitem[{\citenamefont{Kuroki et~al.}(2009)\citenamefont{Kuroki, Usui, Onari,
  Arita, and Aoki}}]{k_kuroki_09}
\bibinfo{author}{\bibfnamefont{K.}~\bibnamefont{Kuroki}},
  \bibinfo{author}{\bibfnamefont{H.}~\bibnamefont{Usui}},
  \bibinfo{author}{\bibfnamefont{S.}~\bibnamefont{Onari}},
  \bibinfo{author}{\bibfnamefont{R.}~\bibnamefont{Arita}}, \bibnamefont{and}
  \bibinfo{author}{\bibfnamefont{H.}~\bibnamefont{Aoki}},
  \bibinfo{journal}{Phys. Rev. B} \textbf{\bibinfo{volume}{79}},
  \bibinfo{pages}{224511} (\bibinfo{year}{2009}).

\bibitem[{\citenamefont{{Paglione} and {Greene}}(2010)}]{j_paglione_10}
\bibinfo{author}{\bibfnamefont{J.}~\bibnamefont{{Paglione}}} \bibnamefont{and}
  \bibinfo{author}{\bibfnamefont{R.~L.} \bibnamefont{{Greene}}},
  \bibinfo{journal}{Nature Physics} \textbf{\bibinfo{volume}{6}},
  \bibinfo{pages}{645} (\bibinfo{year}{2010}).

\bibitem[{\citenamefont{Graser et~al.}(2010)\citenamefont{Graser, Kemper,
  Maier, Cheng, Hirschfeld, and Scalapino}}]{s_graser_10}
\bibinfo{author}{\bibfnamefont{S.}~\bibnamefont{Graser}},
  \bibinfo{author}{\bibfnamefont{A.~F.} \bibnamefont{Kemper}},
  \bibinfo{author}{\bibfnamefont{T.~A.} \bibnamefont{Maier}},
  \bibinfo{author}{\bibfnamefont{H.-P.} \bibnamefont{Cheng}},
  \bibinfo{author}{\bibfnamefont{P.~J.} \bibnamefont{Hirschfeld}},
  \bibnamefont{and} \bibinfo{author}{\bibfnamefont{D.~J.}
  \bibnamefont{Scalapino}}, \bibinfo{journal}{Phys. Rev. B}
  \textbf{\bibinfo{volume}{81}}, \bibinfo{pages}{214503}
  (\bibinfo{year}{2010}).

\bibitem[{\citenamefont{Ortenzi et~al.}(2009)\citenamefont{Ortenzi, Cappelluti,
  Benfatto, and Pietronero}}]{l_ortenzi_09}
\bibinfo{author}{\bibfnamefont{L.}~\bibnamefont{Ortenzi}},
  \bibinfo{author}{\bibfnamefont{E.}~\bibnamefont{Cappelluti}},
  \bibinfo{author}{\bibfnamefont{L.}~\bibnamefont{Benfatto}}, \bibnamefont{and}
  \bibinfo{author}{\bibfnamefont{L.}~\bibnamefont{Pietronero}},
  \bibinfo{journal}{Phys. Rev. Lett.} \textbf{\bibinfo{volume}{103}},
  \bibinfo{pages}{046404} (\bibinfo{year}{2009}).

\bibitem[{\citenamefont{Ikeda et~al.}(2010)\citenamefont{Ikeda, Arita, and
  Kune\v{s}}}]{h_ikeda_10}
\bibinfo{author}{\bibfnamefont{H.}~\bibnamefont{Ikeda}},
  \bibinfo{author}{\bibfnamefont{R.}~\bibnamefont{Arita}}, \bibnamefont{and}
  \bibinfo{author}{\bibfnamefont{J.}~\bibnamefont{Kune\v{s}}},
  \bibinfo{journal}{Phys. Rev. B} \textbf{\bibinfo{volume}{81}},
  \bibinfo{pages}{054502} (\bibinfo{year}{2010}).

\bibitem[{\citenamefont{{Maier} et~al.}(2007)\citenamefont{{Maier}, {Macridin},
  {Jarrell}, and {Scalapino}}}]{t_maier_07}
\bibinfo{author}{\bibfnamefont{T.~A.} \bibnamefont{{Maier}}},
  \bibinfo{author}{\bibfnamefont{A.}~\bibnamefont{{Macridin}}},
  \bibinfo{author}{\bibfnamefont{M.}~\bibnamefont{{Jarrell}}},
  \bibnamefont{and} \bibinfo{author}{\bibfnamefont{D.~J.}
  \bibnamefont{{Scalapino}}}, \bibinfo{journal}{\prb}
  \textbf{\bibinfo{volume}{76}}, \bibinfo{pages}{144516}
  (\bibinfo{year}{2007}).

\bibitem[{\citenamefont{Schulz et~al.}(1992)\citenamefont{Schulz, Allen, and
  Trivedi}}]{Schulz_92}
\bibinfo{author}{\bibfnamefont{W.~W.} \bibnamefont{Schulz}},
  \bibinfo{author}{\bibfnamefont{P.~B.} \bibnamefont{Allen}}, \bibnamefont{and}
  \bibinfo{author}{\bibfnamefont{N.}~\bibnamefont{Trivedi}},
  \bibinfo{journal}{Phys. Rev. B} \textbf{\bibinfo{volume}{45}},
  \bibinfo{pages}{10886} (\bibinfo{year}{1992}).

\bibitem[{\citenamefont{Kim et~al.}(1998)\citenamefont{Kim, Mazin, and
  Singh}}]{Kim_98}
\bibinfo{author}{\bibfnamefont{S.-G.} \bibnamefont{Kim}},
  \bibinfo{author}{\bibfnamefont{I.~I.} \bibnamefont{Mazin}}, \bibnamefont{and}
  \bibinfo{author}{\bibfnamefont{D.~J.} \bibnamefont{Singh}},
  \bibinfo{journal}{Phys. Rev. B} \textbf{\bibinfo{volume}{57}},
  \bibinfo{pages}{6199} (\bibinfo{year}{1998}).

\bibitem[{\citenamefont{Golubov and Mazin}(1997)}]{a_golubov_97}
\bibinfo{author}{\bibfnamefont{A.~A.} \bibnamefont{Golubov}} \bibnamefont{and}
  \bibinfo{author}{\bibfnamefont{I.~I.} \bibnamefont{Mazin}},
  \bibinfo{journal}{Phys. Rev. B} \textbf{\bibinfo{volume}{55}},
  \bibinfo{pages}{15146} (\bibinfo{year}{1997}).

\bibitem[{\citenamefont{{Senga} and {Kontani}}(2008)}]{y_senga_08}
\bibinfo{author}{\bibfnamefont{Y.}~\bibnamefont{{Senga}}} \bibnamefont{and}
  \bibinfo{author}{\bibfnamefont{H.}~\bibnamefont{{Kontani}}},
  \bibinfo{journal}{Journal of the Physical Society of Japan}
  \textbf{\bibinfo{volume}{77}}, \bibinfo{pages}{113710}
  (\bibinfo{year}{2008}).

\bibitem[{\citenamefont{Onari and Kontani}(2009)}]{s_onari_09}
\bibinfo{author}{\bibfnamefont{S.}~\bibnamefont{Onari}} \bibnamefont{and}
  \bibinfo{author}{\bibfnamefont{H.}~\bibnamefont{Kontani}},
  \bibinfo{journal}{Phys. Rev. Lett.} \textbf{\bibinfo{volume}{103}},
  \bibinfo{pages}{177001} (\bibinfo{year}{2009}).

\bibitem[{\citenamefont{Prelov\v{s}ek and Sega}(2010)}]{Prelovsek_10}
\bibinfo{author}{\bibfnamefont{P.}~\bibnamefont{Prelov\v{s}ek}}
  \bibnamefont{and} \bibinfo{author}{\bibfnamefont{I.}~\bibnamefont{Sega}},
  \bibinfo{journal}{Phys. Rev. B} \textbf{\bibinfo{volume}{81}},
  \bibinfo{pages}{115121} (\bibinfo{year}{2010}).

\bibitem[{\citenamefont{Rullier-Albenque
  et~al.}(2010)\citenamefont{Rullier-Albenque, Colson, Forget, Thu\'ery, and
  Poissonnet}}]{Rullier-Albenque_10}
\bibinfo{author}{\bibfnamefont{F.}~\bibnamefont{Rullier-Albenque}},
  \bibinfo{author}{\bibfnamefont{D.}~\bibnamefont{Colson}},
  \bibinfo{author}{\bibfnamefont{A.}~\bibnamefont{Forget}},
  \bibinfo{author}{\bibfnamefont{P.}~\bibnamefont{Thu\'ery}}, \bibnamefont{and}
  \bibinfo{author}{\bibfnamefont{S.}~\bibnamefont{Poissonnet}},
  \bibinfo{journal}{Phys. Rev. B} \textbf{\bibinfo{volume}{81}},
  \bibinfo{pages}{224503} (\bibinfo{year}{2010}).

\bibitem[{\citenamefont{{Devereaux} and {Hackl}}(2007)}]{t_devereaux_rmp}
\bibinfo{author}{\bibfnamefont{T.~P.} \bibnamefont{{Devereaux}}}
  \bibnamefont{and} \bibinfo{author}{\bibfnamefont{R.}~\bibnamefont{{Hackl}}},
  \bibinfo{journal}{Reviews of Modern Physics} \textbf{\bibinfo{volume}{79}},
  \bibinfo{pages}{175} (\bibinfo{year}{2007}).

\bibitem[{\citenamefont{{Devereaux} and {Kampf}}(1999)}]{t_devereaux_99}
\bibinfo{author}{\bibfnamefont{T.~P.} \bibnamefont{{Devereaux}}}
  \bibnamefont{and} \bibinfo{author}{\bibfnamefont{A.~P.}
  \bibnamefont{{Kampf}}}, \bibinfo{journal}{Phys. Rev. B}
  \textbf{\bibinfo{volume}{59}}, \bibinfo{pages}{6411} (\bibinfo{year}{1999}).

\bibitem[{\citenamefont{{Devereaux} et~al.}(1996)\citenamefont{{Devereaux},
  {Virosztek}, and {Zawadowski}}}]{t_devereaux_96}
\bibinfo{author}{\bibfnamefont{T.~P.} \bibnamefont{{Devereaux}}},
  \bibinfo{author}{\bibfnamefont{A.}~\bibnamefont{{Virosztek}}},
  \bibnamefont{and}
  \bibinfo{author}{\bibfnamefont{A.}~\bibnamefont{{Zawadowski}}},
  \bibinfo{journal}{\prb} \textbf{\bibinfo{volume}{54}}, \bibinfo{pages}{12523}
  (\bibinfo{year}{1996}).

\bibitem[{\citenamefont{{Mazin} et~al.}(2010)\citenamefont{{Mazin},
  {Devereaux}, {Analytis}, {Chu}, {Fisher}, {Muschler}, and
  {Hackl}}}]{i_mazin_10a}
\bibinfo{author}{\bibfnamefont{I.~I.} \bibnamefont{{Mazin}}},
  \bibinfo{author}{\bibfnamefont{T.~P.} \bibnamefont{{Devereaux}}},
  \bibinfo{author}{\bibfnamefont{J.~G.} \bibnamefont{{Analytis}}},
  \bibinfo{author}{\bibfnamefont{J.-H.} \bibnamefont{{Chu}}},
  \bibinfo{author}{\bibfnamefont{I.~R.} \bibnamefont{{Fisher}}},
  \bibinfo{author}{\bibfnamefont{B.}~\bibnamefont{{Muschler}}},
  \bibnamefont{and} \bibinfo{author}{\bibfnamefont{R.}~\bibnamefont{{Hackl}}},
  \bibinfo{journal}{\prb} \textbf{\bibinfo{volume}{82}},
  \bibinfo{pages}{180502} (\bibinfo{year}{2010}).

\end{thebibliography}

\end{document}